\documentclass[twocolumn, 10pt, aps, pra, superscriptaddress]{revtex4-1}
\usepackage{amsmath,amssymb,mathtools,xcolor,bbold}
\usepackage{graphicx}
\usepackage{hyperref}
\usepackage[T1]{fontenc}
\usepackage[utf8]{inputenc}
\inputencoding{utf8}

\graphicspath{ {./figures/} }

\begin{document}

\title{Dissipation in adiabatic quantum computers: Lessons from an exactly solvable model}

\author{Maximilian Keck}
\affiliation{NEST, Scuola Normale Superiore and Istituto Nanoscienze-CNR, I-56126 Pisa, Italy}

\author{Simone Montangero}
\affiliation{Theoretische Physik, Universit\"at des Saarlandes, D-66123 Saarbr\"ucken, Germany}
\affiliation{Institute for Complex Quantum Systems \& Center for Integrated
  Quantum Science and Technologies, Universit\"at Ulm, D-89069 Ulm, Germany}

\author{Giuseppe E. Santoro}
\affiliation{SISSA, Via Bonomea 265, I-34151 Trieste, Italy}
\affiliation{ICTP, Strada Costiera 11, I-34151 Trieste, Italy}

\author{Rosario Fazio}
\affiliation{ICTP, Strada Costiera 11, I-34151 Trieste, Italy}
\affiliation{NEST, Scuola Normale Superiore and Istituto Nanoscienze-CNR, I-56126 Pisa, Italy}

\author{Davide Rossini}
\affiliation{Dipartimento di Fisica, Universit\`a di Pisa and INFN, Largo Pontecorvo 3, I-56127 Pisa, Italy}
\affiliation{NEST, Scuola Normale Superiore and Istituto Nanoscienze-CNR, I-56126 Pisa, Italy}

\date{\today}

\begin{abstract}
We introduce and study the adiabatic dynamics of free-fermion models subject to a local Lindblad bath and in the 
presence of a time-dependent Hamiltonian. The merit of these models is that they can be solved exactly,
and will help us to study the interplay between non-adiabatic transitions and dissipation in many-body quantum systems.
After the adiabatic evolution, we evaluate the excess energy (average value of the Hamiltonian) as a measure
of the deviation from reaching the target final ground state.
We compute the excess energy in a variety of different situations, where the nature of the bath and 
the Hamiltonian is modified. We find a robust evidence of the fact that an optimal working time for the quantum annealing 
protocol emerges as a result of the competition between the non-adiabatic effects and the dissipative processes.
We compare these results with matrix-product-operator simulations of an Ising system and show that the phenomenology 
we found applies also for this more realistic case.
\end{abstract}


\maketitle

\section{Introduction}

The recent experimental advances in the field of quantum technologies have drastically enhanced
our capability to control the quantum coherent dynamics of many-body systems in a variety of physical 
systems, ranging from atomic and molecular optics, to trapped ions, and cavity/circuit quantum 
electrodynamics. These progresses have made real the possibility to experimentally realise quantum 
simulators~\cite{NatPhys_2012} as well as the implementation of the first quantum algorithms~\cite{Corcoles_2015,Chow_2014}.

Together with the progresses in implementing quantum gates and concatenate them, i.e., by realising standard circuit 
computation~\cite{NielsenChuang}, recently  adiabatic quantum computation (AQC)~\cite{Fahri_2001}  and quantum 
annealing~\cite{Santoro_2002} have received a tremendous boost thanks to the experiments performed with  
D-Wave machines~\cite{Denchev_2016, Lanting_2014, King_2016, Venturelli_2015}. The  strategy underlying  adiabatic quantum computation~\cite{Fahri_2001, Santoro_2002}
is based on the fact that any quantum algorithm can be formulated in terms of identifying the global minimum (ground 
state) of a given function (Hamiltonian) over a set of many local minima.
On the experimental side, quantum effects were seen to survive on eight~\cite{Johnson_2011, Boixo_2013}, 
sixteen~\cite{Dickson_2013} and even in more than one hundred qubits~\cite{Boixo_2014}.
Whether these machines hold already at present the so called ``quantum supremacy'' or not is still
under debate~\cite{Ronnow_2014, Boixo_2016}. However, it is clearly very important to understand the actual mode
of operation of these adiabatic computers, in order to understand the limit of their performances and to push it forward. 

A key problem in this framework is to understand the role of dissipation and decoherence on adiabatic quantum computers.
This question amounts to understanding the key features that control the adiabatic evolution of a many-body open quantum system.

Let us first state the general problem.
Suppose to be able to follow the quantum dynamics of an appropriate time-dependent Hamiltonian: 
\begin{equation}
  H(t) = [1-f(t)] \, H_{\rm in} + f(t) \, H_{\rm fin}, \quad t \in [t_{\rm in}, t_{\rm fin}],
  \label{eq:QAnnealing}
\end{equation}
$f(t)$ being a generic function of time, with $f(t_{\rm in})=0$ and $f(t_{\rm fin}) = 1$.
The Hamiltonian $H_{\rm in}$ sets the initial condition as its ground state, while the sought solution to the problem
is entailed into the ground state of $H_{\rm fin}$. 
If the control time is much larger than the typical inverse gap between the ground state and the first excited state,
the system will adiabatically follow its instantaneous ground state $|\psi_0(t)\rangle$.
Reaching the ground state of $H_{\rm fin}$ by adiabatic evolution is the way AQC works~\cite{Fahri_2001, Santoro_2002}.
As long as the evolution is unitary, the only source of errors is due to excitations generated by non-adiabatic effects
in the dynamical evolution. The complexity of the adiabatic algorithms is reflected in the scaling of 
the minimum gap with the number of qubits. In general, AQC also requires a special form of $f(t)$ to gain a speedup as compared to classical algorithms, see Ref.~\cite{Roland_2002} for a prominent example of an adiabatic Grover search and Ref.~\cite{Torrentegui_2013} for a review. Moreover, optimal controlled ramps can provide additional speedups~\cite{QOC_2011}.
An alternative protocol, which is more general than Eq.~\eqref{eq:QAnnealing},
would incorporate an extra (possibly non-linear) term, such as $f(t)[1-f(t)]H_{\rm E}$,
with $H_E$ being properly chosen Hamiltonian. This may have beneficial
effects on the minimum gap of the system, and therefore greatly enhance the AQC performance~\cite{non-linearAQC}.

It is however clear that, especially for long annealing times, another important source of defects is related to incoherent
fluctuations induced by finite temperature, or more in general by the unavoidable coupling of the system to some
external environment. In this case the quantum state of the system will be mixed, described by a density matrix $\rho(t)$,
satisfying a dynamical equation of the form:
\begin{equation}
  \frac{\partial \rho}{\partial t} = - \frac{i}{\hbar} [H(t), \rho] + \mathbb{D}[\rho] .
  \label{eq:master}
\end{equation}
The first term on the right-hand side describes the coherent unitary time evolution, which is ruled by a many-body time-varying 
Hamiltonian $H(t)$, according to the quantum annealing protocol~\eqref{eq:QAnnealing}.
The second term on the right-hand side of Eq.~\eqref{eq:master} accounts for the coupling to the environment,
and its form will depend on the nature of noise and dissipation as well as the form of the coupling
between the many-body system and the external bath. The dissipator is a completely positive, trace preserving map, that -- in general -- drives the system into a fixed point, that is a steady state or a steady-state manifold (if the fixed point is not unique). As such, it is inducing a decay in the system towards the steady state.
Note that the differential form of Eq.~\eqref{eq:master} generally requires a Markovian bath.
Understanding the 
effect of dissipation on AQC amounts to quantifying, in some way, how the state of the system deviates from the final
ground state, because of the presence of the extra term $\mathbb{D}[\rho]$ in Eq.~\eqref{eq:master}.

Does the presence of the environment facilitate the reaching of the final ground state or is it detrimental for AQC?
It is clear that there cannot be a unique answer to this question: the deviations from the unitary case
may depend strongly on the form of $\mathbb{D}[\rho]$ in relation to the type of evolution imposed by $H(t)$.
This variety of possible answers is reflected in the wide spectrum of cases already considered in the literature.
Within the plethora of possible scenarios, it is however important to establish some general trends that may serve 
as guidelines in going deeper in this formidable problem. 

This type of analysis was first performed~\cite{Patane_2008} in the context of the Kibble-Zurek (KZ) mechanism for defect 
formation~\cite{Kibble, Zurek}. The KZ mechanism is related to the 
fact that, when crossing a gapless critical point of $H(t)$, no matter how slow the variation of the Hamiltonian in time is, the 
adiabatic theorem is violated and a finite density of defects will be produced. More than thirty years ago, Kibble put 
forward a scaling argument aimed at predicting the size and the number of such defects~\cite{Kibble}, while the mechanism yielding the correct scaling was found later by Zurek~\cite{Zurek},
roughly dividing the dynamics in either adiabatic or impulsive, according to the distance from the critical point.
The KZ mechanism has been tested in a variety of quantum toy models at zero temperature, including ordered and disordered systems,
as well as for crossing isolated or extended critical regions (see, e.g., Ref.~\cite{Dziarmaga_rev} for a review).
When the annealing velocity is progressively increased, a crossover behaviour sets in between the KZ scaling
and the generation of excitations due to faster quenches, where the dynamics can be described
by the underlying classical model~\cite{Silvi_2016}.

The proliferation of defects due to Landau-Zener transitions is intimately related
to the occurrence of errors in the AQC. While, in the unitary case, the number of defects decreases on 
increasing annealing time, the environment will be dominant for long annealing times.
In this regime, one expects a defect formation which is almost independent on the annealing protocol.
This picture was confirmed and detailed in Ref.~\cite{Patane_2008}, where the scaling in the crossover between
the KZ-dominated regime and the environment-dominated regime was also found. In the presence of spatially correlated 
noise, additional intermediate regimes emerge due to the comparison of the correlation length of the noise
and the correlation length of the system~\cite{Nalbach_2015}.

The role of temperature and external noise was further considered in the context of AQC in several papers,
showing that in some cases it may be beneficial in reaching the target ground state.
Work has been done on the comparison between AQC and classical approaches using thermal hopping~\cite{Boixo_2016b},
on the use of quantum diffusion, showing better performance than closed-system quantum annealing~\cite{Smelyanskiy_2017}
as well as on the crucial role of noise-induced thermalisation in AQC that can outperform
simulated annealing~\cite{Kechedzhi_2016}. Further, the relation of thermally-assisted tunneling to quantum Monte Carlo has been studied in Ref.~\cite{Jiang_2017}. In addition to this, the effect of noise of a thermal environment~\cite{Albash_2015} and decoherence~\cite{Albash_2012} have been studied as well.

It should be kept in mind that the adiabatic dynamics of dissipative many-body system is linked to the understanding
of the Landau-Zener problem of a two-level system coupled to an environment, an extensive studied problem
in many different areas~\cite{Ao_1989, Wubs_2006, Ashhab_2006, Zueco_2012, Javanbakht_2015, Arceci_2017}.

The problem of describing the adiabatic dynamics of a many-body open quantum system is a formidable problem and approximations
are necessary. It is however of fundamental importance to have some non-trivial examples where the outcomes of the analysis
are not hampered by any approximation. The aim of this work is to present some simple, yet non-trivial, examples
where the adiabatic dynamics can be analysed in full details. Most importantly, the phenomenology that we will extract
is related to the dynamics of certain spin systems that are very close to the relevant implementations of AQC.
This means that the exactly solvable models that we consider here may be used as a very useful 
benchmark to test important approximations in more complex cases. 

The bath we will deal with is Markovian. This means that the dissipative term in Eq.~\eqref{eq:master} can be written
in the Lindblad form 
\begin{equation}
  \mathbb{D} [\rho] = \sum_n \kappa_n \Big( L_n \rho L^\dagger_n - \tfrac12 \{ \rho, L^\dagger_n L_n \} \Big) ,
  \label{eq:Lindbladian}
\end{equation}
where $L_n$ are suitable local Lindblad operators that describe the environment (to be defined later)
and $\kappa_n$ are the corresponding couplings, which have to be positive for a Markovian Lindblad master equation. The choice of local Lindblad operators that we are going to study does not lead to a thermal state in the steady state. We will dwell in more details on this point later.

We will consider quadratic fermionic models whose Lindblad dynamics
can be worked out analytically~\cite{Prosen_2008, Eisler_2011, Horstmann_2013}, for different types
of local system-bath coupling. The dynamics of this class of models can be studied exactly, and it will help us 
to clarify several features of the interplay between non-adiabatic effects and incoherent transitions due to the external bath. 
In order to understand to which extent the results we find can be applied to more realistic cases, later we will consider
a spin-$1/2$ one-dimensional Ising model. In such case, for the incoherent spin decay/pumping,
the master equation cannot be mapped into local fermionic operators,
therefore we will resort to a numerical study based on a matrix-product-operator (MPO)
representation of the density matrix~\cite{Verstraete_2004, Zwolak_2004}. As we will discuss in more details in the rest
of the paper, the overall phenomenology remains unchanged, thus reinforcing the fact that the exactly solvable models
introduced here can be very useful benchmarks.
We should remind that, in general, the thermodynamic properties of low-dimensional systems
can be strongly affected by the dimensionality: for example, thermal fluctuations wash out quantum fluctuations
of finite-temperature systems in one dimension, but not anymore in two dimensions.
Furthermore, methods that are very powerful in one dimension, as the Jordan-Wigner transformation, are not applicable
in higher dimensions. However, in the special case of free fermions, the system's thermodynamics 
is not expected to change with the dimensionality.

In all the situations we have addressed, we find a robust evidence of the fact that an optimal working time 
for the quantum annealing protocol emerges as a result of the competition between
the non-adiabatic effects and the dissipative processes.
There is an optimal time for which the final state is closest to the true final ground state.
The scaling of such optimal time (and that of the corresponding generated defects) can be accurately
predicted by assuming that the number of defects produced during the time evolution
is a sum of the two contributions due to non-adiabaticity and dissipation/decoherence.

For larger working times it may happen that, depending on the type of system-bath coupling,
an overshooting point sets in, where the density of the generated defects is larger than that for
an infinitely slow annealing, which would adiabatically drive the system through the instantaneous steady state.
While this kind of behaviour cannot appear in the unitary scenario, where the defects production
is monotonic non-increasing with the annealing speed, in the system-bath scenario this can emerge even for small systems,
being eventually related to the spectral structure of the Liouvillian and not necessarily to many-body characteristics.

The paper is organised as follows. In Sec.~\ref{sec:Model} we define the Hamiltonian models,
the various dissipation schemes, and the annealing problem under investigation.
We then study the departure from the instantaneous ground state,
in the presence of dissipative processes which may cause decay or dephasing,
both for a translationally invariant free-fermion model~\ref{sec:TraslInv},
and for an Ising spin chain (Sec.~\ref{sec:Ising}).
We end with a discussion of our findings and with the concluding remarks in Sec.~\ref{sec:Conclusions}.
Technical details on the calculations for the quadratic fermionic model are provided in the Appendix.

\section{Adiabatic dynamics with local dissipation: From Ising systems to fermion chains}
\label{sec:Model}

One of the simplest (and exactly solvable) models exhibiting a quantum phase transition
is the one-dimensional Ising chain~\cite{Sachdev}. This is defined by the Hamiltonian
\begin{equation}
  \label{eq:Ham_Ising}
  H(t) = - J \sum_{n} \sigma^{x}_{n}\sigma^{x}_{n+1} - \Gamma(t) \sum_{n} \sigma^{z}_{n},
\end{equation}
where $\sigma^{\alpha}_{n} (\alpha = x,y,z)$ are the spin-$1/2$ Pauli operators
for the $n$-th spin of the chain, 
while $J$ and $\Gamma(t)$ are respectively the coupling strength between neighbouring spins
and the transverse magnetic field.
We assume periodic boundary conditions, in such a way as
to preserve translational invariance of the model.
We will also work in units of $\hbar = 1$ and set the energy scale by fixing $J=1$.

It is possible to realise quantum annealing in the Ising model by tuning
the parameter $\Gamma(t)$ in time according to a linear ramping, as for example:
\begin{equation}
  \Gamma(t) = - t / \tau, \quad \text{for} \ \ t \in (-\infty,0] ,
    \label{eq:annealing}
\end{equation}
where $\tau$ is related to the ramping speed.
The choice~\eqref{eq:annealing} ensures that, during the annealing procedure,
the system will encounter a critical point in which the ground-state energy gap closes
and defects will start to appear.
Duality arguments~\cite{Fisher_1995} show that the phase transition occurs at $\Gamma_{\rm c}=1$.
The system is driven from a paramagnetic phase, where all the spins are aligned
along the field direction $z$ [i.e., the ground state of the initial Hamiltonian
$H(t_{\rm in}) \propto -\sum_n \sigma^z_n$, since $\Gamma(t_{\rm in}) = +\infty$],
to a doubly-degenerate ferromagnetic phase, where the spins are all
pointing along the coupling direction $x$ [i.e., the ground state of the final Hamiltonian
$H(t_{\rm fin}) = -\sum_n \sigma^x_n \sigma^x_{n+1}$, since $\Gamma(t_{\rm fin}) = 0$].

Several papers have already addressed the KZ scaling of defects with $\tau$ in the paradigmatic
Ising model, both for the clean system~\cite{Zurek_2005, Dziarmaga_2005} and
for the disordered system~\cite{Dziarmaga_2006, Caneva_2007}.
Here we are going to add the effect of the coupling to an external environment,
modelled through a master equation of the form in Eq.~\eqref{eq:master}.

To retain analytic solvability of the full open-system problem~\cite{Prosen_2008, Eisler_2011},
we will however start from a mapped version of the Ising chain~\eqref{eq:Ham_Ising}
into a free-fermion model.
The latter can be achieved by employing a Jordan-Wigner transformation (JWT),
which maps the spin operators in terms of spinless fermions:
\begin{equation}
  \sigma_n^- = \exp \Big( i \pi \sum_{m<n} c^\dagger_m c_m \Big) c_n ,
  \label{eq:JWT}
\end{equation}
where $\sigma_n^\pm = \tfrac12 (\sigma_n^x \pm i \sigma_n^y)$, while $c_n$ ($c_n^\dagger$) denotes
the fermionic annihilation (creation) operator on site $n$,
obeying the anticommutation relations $\{ c^\dagger_m ,c_n \} = \delta_{m,n}$,
$\{ c_m, c_n \} = 0$.
The resulting Hamiltonian for an Ising chain of length $L$ is quadratic in such operators and reads: 
\begin{equation}
  \label{eq:Ham_quadratic}
  H(t) = \sum_{m,n} \Big[ c^{\dagger}_{m} \, A_{m,n}(t) \, c_{n}
  + \tfrac{1}{2} (c^{\dagger}_{m} \, B_{m,n} \, c^\dagger_{n} + \text{H.c.}) \Big],
\end{equation}
where $A, B$ respectively are a symmetric and an antisymmetric $L \times L$ matrix
whose sole non-zero elements are $A_{n,n}(t) = - \Gamma(t), A_{n,n+1} = A_{n+1,n} = - J/2 ,
B_{n,n+1} = - B_{n+1,n} = - J/2$.
To enforce periodic boundary conditions, the following matrix elements are non-zero as well:
$A_{1,L} = A_{L,1} = (-1)^{N_{\rm F}} J/2$ and $B_{L,1} = - B_{1,L} = (-1)^{N_{\rm F}} J/2$,
where $(-1)^{N_{\rm F}}$ denotes the parity of the number of fermions
$N_{\rm F} = \sum_{n} c^{\dagger}_n c_n$, which commutes with $H$.
The Hamiltonian~\eqref{eq:Ham_quadratic} can be exactly diagonalized 
using a Bogolibuov transformation~\cite{LSM_1961, Young_1997}.

In the following we will completely relax the requirement on fermion-parity dependent 
boundary conditions, and we simply assume that anti-periodic boundary conditions for the
fermions are always enforced. 
This assumption is perfectly justified for a purely coherent evolution, where the fermion parity 
is conserved and the initial ground state has an even number of fermions. 
The reason for enforcing the requirement even when considering the open-system adiabatic dynamics 
is that, as explained below, our Lindblad operators change the fermion parity, and it would be impossible 
to solve the problem by using a parity-dependent boundary conditions. 
More in detail, we will study below the adiabatic dynamics of $H(t)$ under the action of three 
different types of memoryless {\em local} environments.
We model them in such a way that the Lindbladian~\eqref{eq:Lindbladian} is a sum of terms that act 
uniformly ($\kappa_n = \kappa, \; \forall n$) on each site $n$ of the chain:
\begin{align}
  i)   \;\; & L_{n}^{\rm (1)} = c^{\dagger}_{n}         && \mbox{pumping mechanism} , \label{eq:L_pump}\\
  ii)  \;\; & L_{n}^{\rm (2)} = c_{n}                   && \mbox{decaying mechanism} , \label{eq:L_decay}\\
  iii) \;\; & L_{n}^{\rm (3)} = c^{\dagger}_{n} c_{n}   && \mbox{dephasing mechanism} . \label{eq:L_dephas}
\end{align}
Note that while the fermionic dephasing environment~\eqref{eq:L_dephas} can be directly mapped into a dephasing
environment of spins, since there is direct local mapping $c^{\dagger}_{n} c_{n} \to \tfrac12 (\sigma^{z}_{n}+1)$,
the same is not true for the decay and pumping. Indeed, when mapping from a spin operator $\sigma^{-}_{n}$
($\sigma^+_n$) to a fermionic operator $c_{n}$ ($c^\dagger_n$), the Jordan-Wigner
transformation~\eqref{eq:JWT} includes a non-local operator (string).
We should stress that the naming of these Lindblad operators has been chosen with respect to their action
on a system with a local, diagonal Hamiltonian, and not with respect to their actual effect on the system that we study.
The choice of these specific Lindblad operators is motivated by the possibility to study, in an essentially exact way,
the competition between the unitary dynamics and dissipative effects, focusing on features that do not depend
qualitatively on the form of the coupling to the environment.

In the next sections we will first address analytically the fermionic model~\eqref{eq:Ham_quadratic}
with dissipation provided by Lindblad terms as in~\eqref{eq:L_decay}-\eqref{eq:L_dephas},
without making any reference to the mapping with spins.
The effect of the non-local part of the JWT will be only discussed in Sec.~\ref{sec:Ising},
where we will study numerically the Ising model~\eqref{eq:Ham_Ising} with a spin-decay mechanism
provided by $L_n^{(1a)} = \sigma_n^-$ that generalises Eq.~\eqref{eq:L_decay} to spins.
We restate that this choice of local Lindblad operators does not lead to thermalisation in the steady state,
as for this purpose non-local terms would be required. However they cover a special interest since in several experimental
implementations, such as circuit-QED, cold-atom settings or trapped ions, this kind of local damping is the relevant one.

In order to quantify the loss of adiabaticity during the annealing protocol,
originating both from the closure of the Hamiltonian gap and from the dissipative processes,
we are going to study the excess energy $\varepsilon$ per site, at the end of the annealing.
The excess energy at a given time $t$ expresses the difference between
the instantaneous energy during the annealing, $E(t) = {\rm Tr} \big[ H(t) \, \rho(t) \big]$,
where $\rho(t)$ is the solution of the master equation~\eqref{eq:master} at time $t$,
and the ground-state energy
$E_0(t) = {\rm Tr} \big[ H(t) |\psi_0(t)\rangle \langle \psi_0(t)| \big]$
of the instantaneous Hamiltonian system described by $H(t)$:
\begin{equation}
  \label{eq:excess_energy}
  \varepsilon(t) = \tfrac{1}{L} \Big\{ {\rm Tr} [ H(t) \, \rho(t) ]
  - \langle \psi_0(t)|H(t)|\psi_0(t) \rangle \Big\}.
\end{equation}
Using the aforementioned Bogoliubov transformation, the second term of Eq.~\eqref{eq:excess_energy}
can be computed straightforwardly, while the first term $E(t)$ is non trivial (see the Appendix).
For the Ising spin system we will resort to a fully numerical MPO approach.
We point out that a related quantity of interest is the density of defects
$\mathcal{N} \equiv \tfrac{1}{2L} \sum_{n} \langle 1-\sigma^{x}_{n}\sigma^{x}_{n+1} \rangle$,
which, in the case of ordered chains and at the end of the annealing, is equivalent to the excess energy
$\varepsilon(0)$, apart from trivial constants.

\section{Free-fermionic system}
\label{sec:TraslInv}

We first analyse a fermionic system described by the Hamiltonian in
Eq.~\eqref{eq:Ham_quadratic}, where anti-periodic boundary conditions are imposed.
A Fourier transform drastically helps in the diagonalisation of the unitary problem,
since the different momentum modes decouple (see App.~\ref{app:TraslInv_U}).
Note that for the sake of simplicity, here we only consider one-dimensional systems,
but our analysis of fermions can be easily extended to larger dimensionalities, 
since a larger dimension will affect calculations only by changing the Brillouin zone.

Let us concentrate on the case in which each lattice site is coupled to some external
bath through a pumping mechanism as in Eq.~\eqref{eq:L_pump}.
The master equation~\eqref{eq:master} during the annealing protocol can be
easily integrated via a straightforward generalisation of the time-dependent
Bogoliubov method already employed by Dziarmaga~\cite{Dziarmaga_2005},
as detailed in App.~\ref{app:TraslInv_Decay}.
The crucial point resides in the fact that, as for the Hamiltonian,
the dissipative part of the Lindbladian with $L^{(1)}_n=c^{\dagger}_n$ does not mix the various modes
at different momenta, once a Fourier transform has been employed.
As a consequence, the density matrix at time $t$ factorizes
into different contributions for the various modes:
\begin{equation}
  \rho(t) = \bigotimes_k \rho_k(t).
  \label{eq:RhoT_k}
\end{equation}
The relevant Hilbert space for each positive momentum $k$ has dimension 4, and thus
the Liouvillian dynamics can be easily followed inside it.
We recall that, for the unitary Schr\"odinger dynamics, a further decomposition into independent
$2 \times 2$ problems was possible, due to the additional conservation
of the fermionic parity (which is now violated by the dissipative decaying terms).

\begin{figure}[t]
  \centering
  \includegraphics[width=\columnwidth]{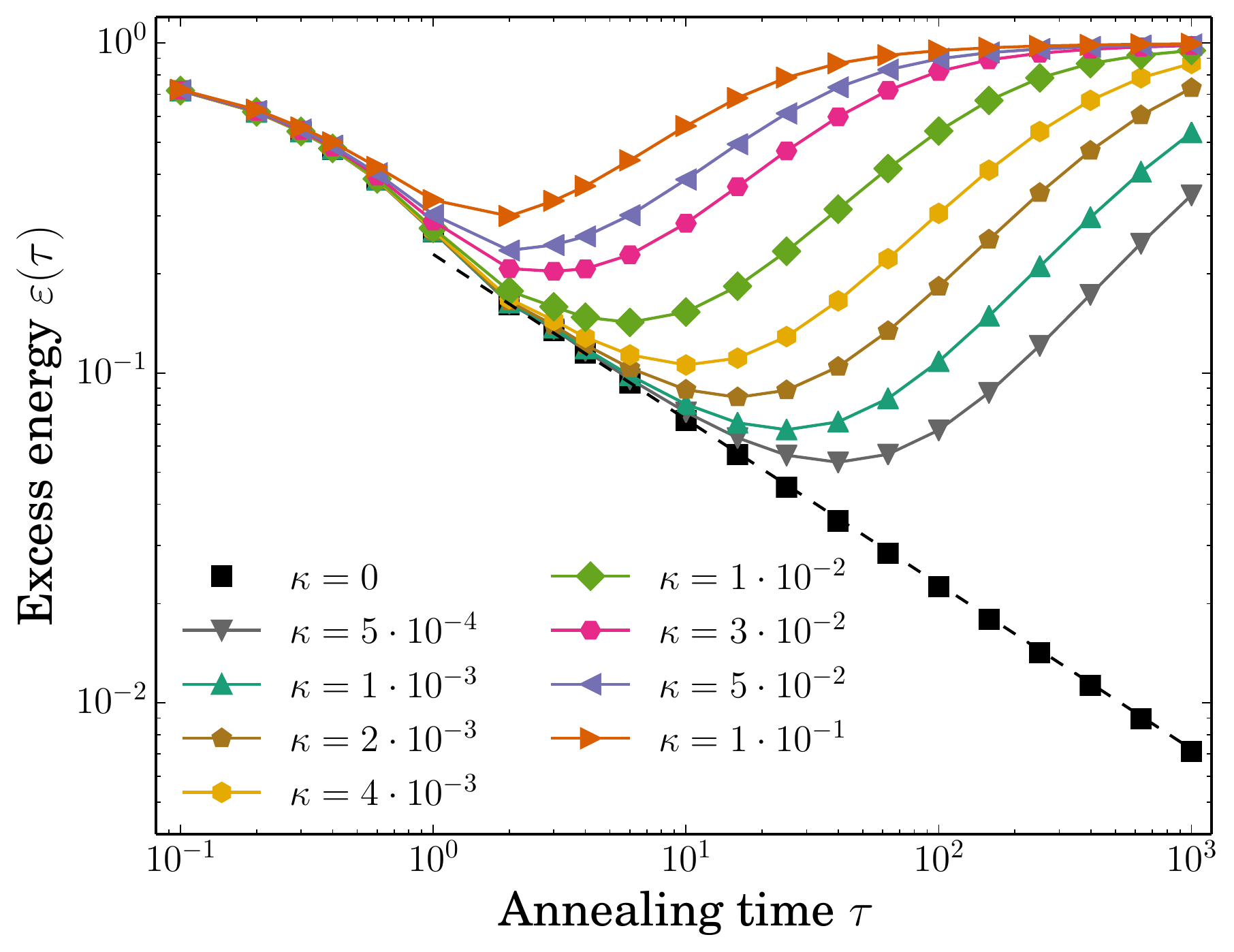}
  \caption{Final excess energy as a function of the annealing time, for the free-fermion
    model~\eqref{eq:Ham_quadratic} coupled to an environment which induces
    a pumping mechanism, as in Eq.~\eqref{eq:L_pump}: $L_n^{(1)} = c^{\dagger}_n$.
    The various data sets denote different values of the dissipative coupling $\kappa$,
    as listed in the legend. Here we simulated the annealing protocol of
    Eq.~\eqref{eq:annealing} for chains of $L=10^3$ sites.
    Black squares denote data for $\kappa=0$, which obey a power-law behaviour
    for $\tau > 1$ with the KZ scaling exponent $\gamma = 0.5$ (dashed line).}
  \label{fig:ExcT_FreeF-Pump}
\end{figure}

The excess energy per site $\varepsilon$ during the annealing protocol
is thus obtained via a numerical integration of the linearized Liouville
equations for each $k$ mode~\eqref{eq:vectorMEQ}.
For numerical convenience, we restricted the initial point of
the annealing procedure~\eqref{eq:annealing} to $t_{\rm in} = -5 \tau$, and checked
that the results are not appreciably affected by this choice~\cite{Caneva_2007}.
We studied systems up to $L=10^3$ sites and annealing times up to $\tau = 10^3$;
a fourth-order Runge-Kutta integration procedure with time step $dt = 10^{-2}$
has been employed.

Fig.~\ref{fig:ExcT_FreeF-Pump} shows the behaviour of the excess energy at the end of the annealing protocol,
$\varepsilon(0)$, for various values of the dissipation strength $\kappa$, as a function of the annealing time $\tau$.
In the absence of dissipation ($\kappa=0$), we recover the Kibble-Zurek (KZ)
scaling~\cite{Zurek_2005, Dziarmaga_2005}
\begin{equation}
  \varepsilon(\tau) \sim 1/ \tau^\gamma \quad \mbox{with} \quad \gamma=1/2 ,
\end{equation}
which can be obtained by the knowledge of the Ising critical exponents associated
to the phase transition at $\Gamma_{\rm c}=1$ across which the system is driven.
A finite dissipation $\kappa>0$ induces a competition between the KZ mechanism of
defect generation due to the crossing of a gapless point (which is progressively reduced,
with increasing annealing time $\tau$), and the production of defects
generated by the incoherent driving itself.
Such competition clearly emerges in Fig.~\ref{fig:ExcT_FreeF-Pump} as a non-monotonic behaviour, 
which generates an optimal working point for the annealing procedure in the presence of dissipation. 

\begin{figure}[!t]
  \centering
  \includegraphics[width=\columnwidth]{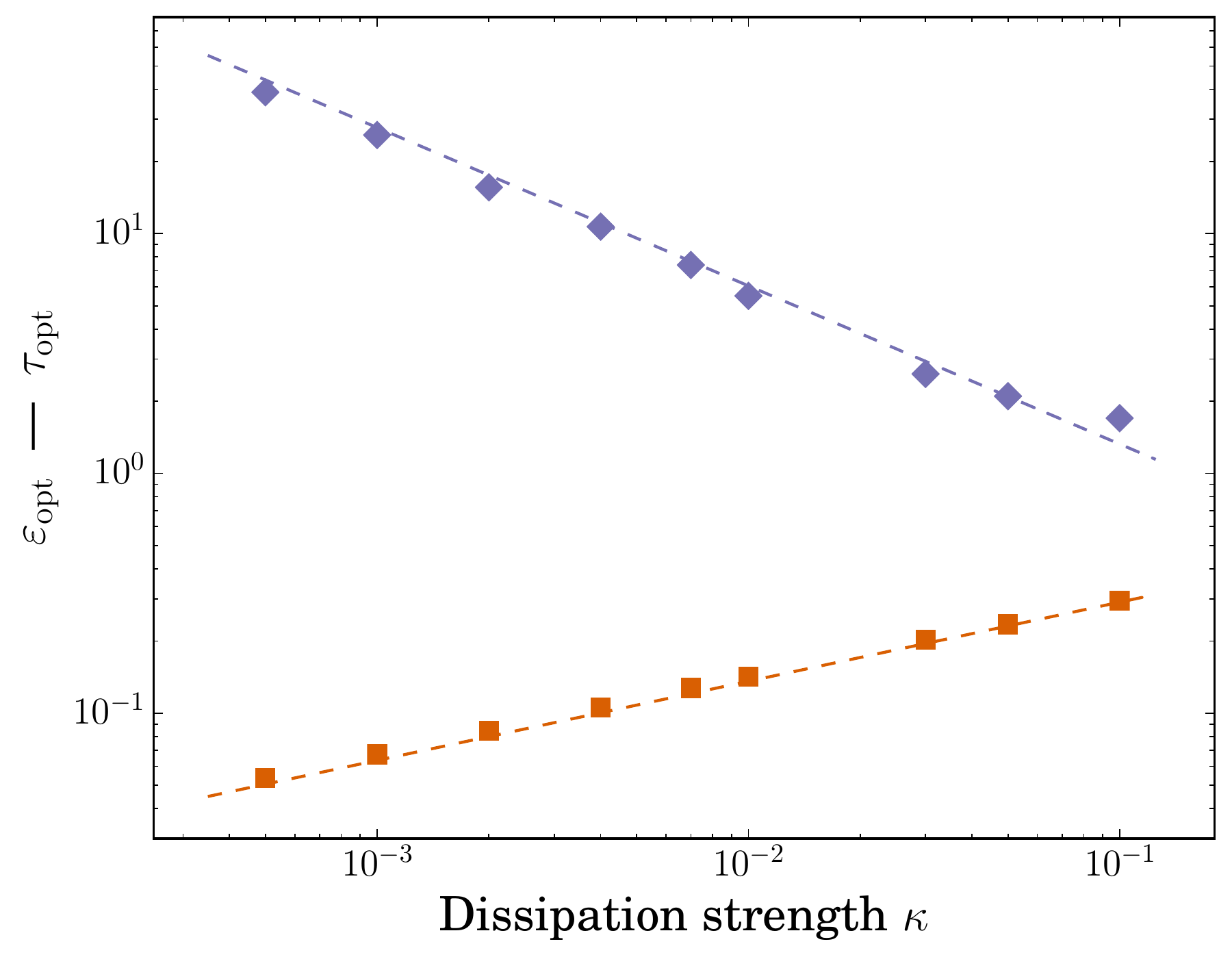}
  \caption{Optimal excess energy $\varepsilon_{\rm opt}$ (orange squares) and corresponding
    annealing time $\tau_{\rm opt}$ (violet diamonds), as a function of the dissipation strength $\kappa$.
    Numerical data (symbols) are obtained using the same parameters as in Fig.~\ref{fig:ExcT_FreeF-Pump},
    and nicely follow a power-law behaviour (dashed lines) with slope $1/3$ and $-2/3$, respectively.}
  \label{fig:OptT_FreeF-Pump}
\end{figure}

Let us now have a closer look at the non-monotonicity, and focus on
the optimal (minimal) value $\varepsilon_{\rm opt}$ reached by the excess energy,
and on the corresponding annealing time $\tau_{\rm opt}$.
Figure~\ref{fig:OptT_FreeF-Pump} displays how such quantities depend on $\kappa$.
Our numerical data nicely agree with a power-law behaviour over more than
two decades of $\kappa$ values, such that $\varepsilon_{\rm opt} \sim \kappa^{1/3}$
and $\tau_{\rm opt} \sim \kappa^{-2/3}$.
Below we show that this behaviour can be easily predicted by assuming that the KZ production
of defects is totally independent of that generated by the dissipation.
The above mentioned competition is thus explained in terms of an incoherent
summation of the two (independent) contributions.

\subsection{Scaling of the optimal point}
\label{sec:ScalOpt}

We start from the observation that, after the annealing procedure, the final state
of the closed system can be easily written as a Bogoliubov state where excitations
are provided by pairs of quasiparticles with equal and opposite momenta~\cite{Dziarmaga_2005}:
\begin{equation}
  |\psi(t_{\rm fin})\rangle = \prod_{k>0} (\alpha_{k} + \beta_{k} \gamma^{\dagger}_{k}\gamma^{\dagger}_{-k}) |0\rangle.
\end{equation}
Here $|0\rangle$ indicates the Bogoliubov vacuum corresponding to the final ground state of $H(0)$,
$\alpha_k$ and $\beta_k$ are complex amplitudes,
while the momentum $k$ can take $L/2$ positive values from 0 to $\pi$
(see App.~\ref{app:TraslInv_U} for details).

In the dissipative case, we will not only have those doubly excited states
$|1_k, 1_{-k}\rangle = \gamma_k^\dagger \gamma_{-k}^\dagger |0\rangle$,
but also singly excited states such as $|1_{k}\rangle = \gamma_k^\dagger |0\rangle$ and
$|1_{-k}\rangle = \gamma_{-k}^\dagger |0\rangle$, which represent further sources of defects.
Indeed, by using the Bogoliubov transformation, we can rotate the Master equation in this frame.
This allows us to write down the dynamical equation for $\langle1_{k}|\rho_{k}|1_{k}\rangle$. We find
\begin{equation}
  \tfrac{\mathrm{d}}{\mathrm{d}t} \langle1_{k}|\rho_{k}|1_{k}\rangle
  = \kappa \, \langle0|\rho_{k}|0\rangle \, f(\Gamma,k),
\end{equation}
with
\begin{equation}
\label{eq:scaling_function}
  [f(\Gamma,k)]^{-1} = 1 \! + \!
  \bigg( \dfrac{\Gamma - \cos k + \sqrt{1 \! + \! \Gamma^2 \! - \! 2\Gamma \cos k}}{\sin^2 k} \bigg)^2
\end{equation}
for the specific choice of $L_{n} = c^{\dagger}_{n}$. In the adiabatic regime where the KZ scaling argument holds and for small dissipation,
the density of defects is much smaller than $1$, so that $\langle0|\rho_{k}|0\rangle$ can be approximated
by its initial value $1$. Note that, since the density of defects $\mathcal{N}$
is written as $\mathcal{N} = \sum_{k} \gamma^{\dagger}_{k}\gamma_{k}$ in the Bogoliubov basis,
excitations of the form $|1_{k}\rangle$ only contribute to the positive values of $k$,
while excitations due to coherent dynamics $|1_{k} 1_{-k}\rangle$ contribute to both, $k$ and $-k$.
Following this, the incoherent part of density of defects can be estimated according to
\begin{equation}
  \mathcal{N}_{\text{inc}} = \dfrac{\kappa}{L} \sum_{k>0}\int_{-\infty}^{0} \mathrm{d}t\ f[\Gamma(t),k]
  = \dfrac{1}{2} \kappa \, \tau,
\end{equation}
where the last equality has been obtained after a change of variables from $t$ to $\Gamma(t) = -t/\tau$,
and observing that the summation over $k>0$ after the integral over $\Gamma$ yields a constant factor $L/2$.

Assuming now that the mechanisms of defect generation due to KZ and due to dissipation are unrelated~\cite{Dutta_2016},
we have:
\begin{equation}
  \mathcal{N} \sim \mathcal{N}_{\text{KZ}} + \mathcal{N}_\text{inc}
  = \dfrac{1}{2\pi \sqrt{2}} \, \tau^{-1/2} + \dfrac{1}{2} \kappa \, \tau.
\end{equation}
From this expression for the total density of defects, the optimal annealing time
minimizing the defects production can be thus estimated by the condition
$\partial_\tau \mathcal{N}(\tau) |_{\tau_{\rm opt}} = 0$.
A direct calculation gives
\begin{equation}
  \tau_{\rm opt} = \left( \dfrac{1}{2\pi \sqrt{2}}\right)^{2/3} \kappa^{-2/3},
  \label{eq:TauOpt}
\end{equation}
with a corresponding density of defects
\begin{equation}
  \mathcal{N}_{\rm opt} = \mathcal{N}(\tau_{\rm opt}) = \dfrac{3}{2} \left( \dfrac{1}{2\pi \sqrt{2}}\right)^{2/3} \kappa^{1/3} .
  \label{eq:NOpt}
\end{equation}

The predictions given by these equations are in nice agreement with our numerical data
shown in Fig.~\ref{fig:OptT_FreeF-Pump}, keeping in mind that $\varepsilon = 2 \mathcal{N}$.

\begin{figure}[!t]
  \centering
  \includegraphics[width=\columnwidth]{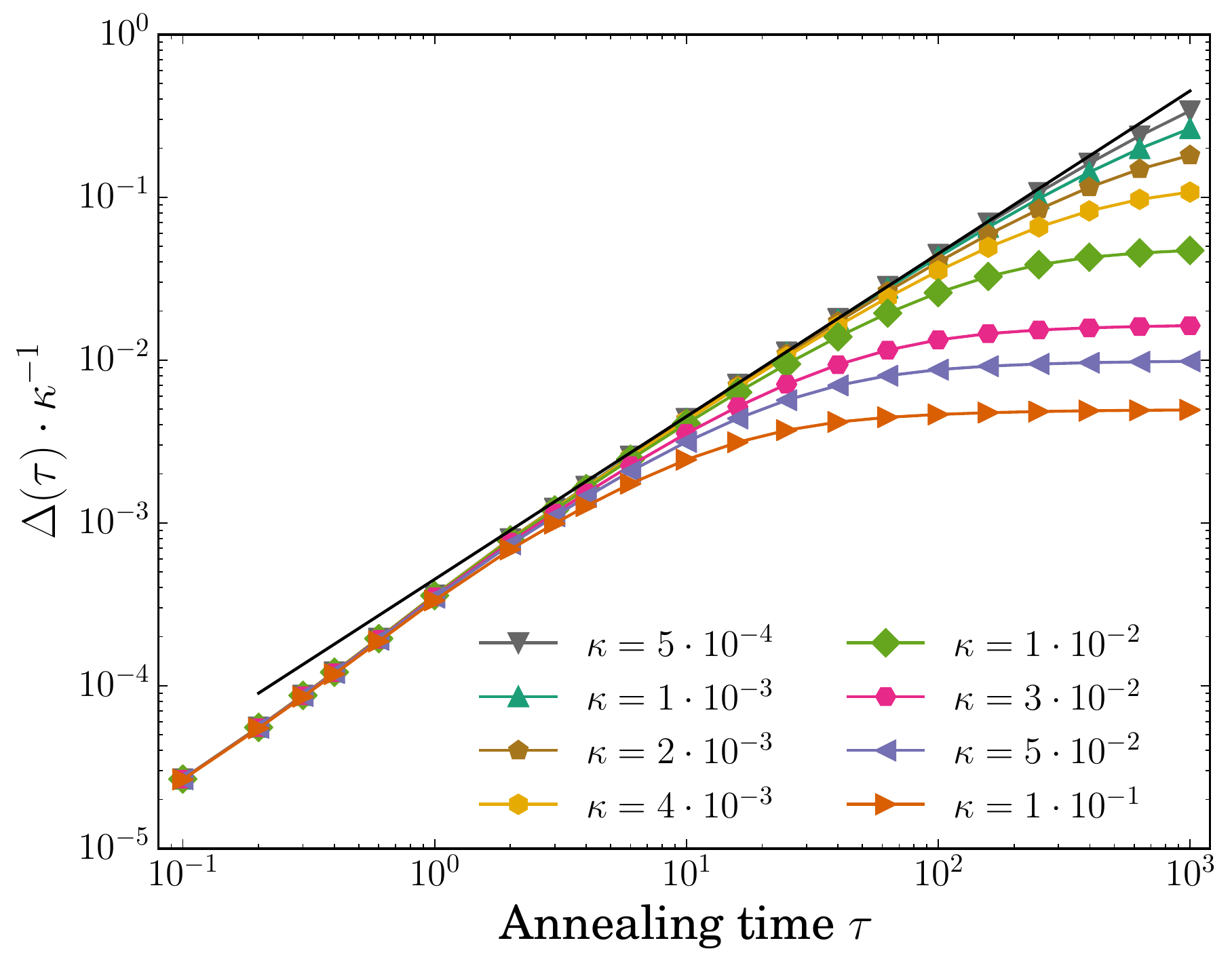}
  \caption{Final excess-energy difference $\Delta$ as a function of $\tau$, once rescaled by $\kappa$.
    The various data sets stand for different values of $\kappa$, and correspond to those
    of Fig.~\ref{fig:ExcT_FreeF-Pump}, where the same colour code has been used.
    A straight line indicating the linear scaling in the annealing time $\tau$ is shown in black.}
  \label{fig:Rscl_FreeF-Pump}
\end{figure}

To further highlight the role of the dissipation during the annealing procedure, we have also analysed
the excess energy at the end of the annealing, after subtracting the corresponding excess energy
in the absence of dissipation:
\begin{equation}
  \Delta(\tau) = \varepsilon(\kappa,\tau) - \varepsilon(\kappa=0,\tau).
  \label{eq:Delta_def}
\end{equation}
Note that, in order to properly define the quantity $\Delta$, we have manifested in Eq.~\eqref{eq:Delta_def}
the $\kappa$-dependence of $\varepsilon$.
After rescaling such quantity as $\Delta(\tau) \to \Delta(\tau) / \kappa$,
we observe a fairly good data collapse with $\tau$, as plotted in Fig.~\ref{fig:Rscl_FreeF-Pump}.
In addition, our data obey a linear scaling as a function of the annealing time except for deviations
induced by bigger values of $\kappa$ (rather than by longer annealing times $\tau$)
in the regime where the excess energy is nearly saturated to its maximal value (see also Fig.~\ref{fig:ExcT_FreeF-Pump}).
We have checked that the behaviour of $\varepsilon(\tau) - \varepsilon(+\infty)$ towards saturation
decays with a power law as $\sim \tau^{-1}$, which is in accordance with Ref.~\cite{CamposVenuti}.

The observations made above point toward a substantial independence of the role played
by the dissipation, with respect to the KZ mechanism. The incoherent coupling to the external bath
acts uniformly and irrespective of the adiabaticity condition ruled by the ground-state energy gap.

\subsection{Interplay between pumping and decay}

\begin{figure}[!t]
  \centering
  \includegraphics[width=\columnwidth]{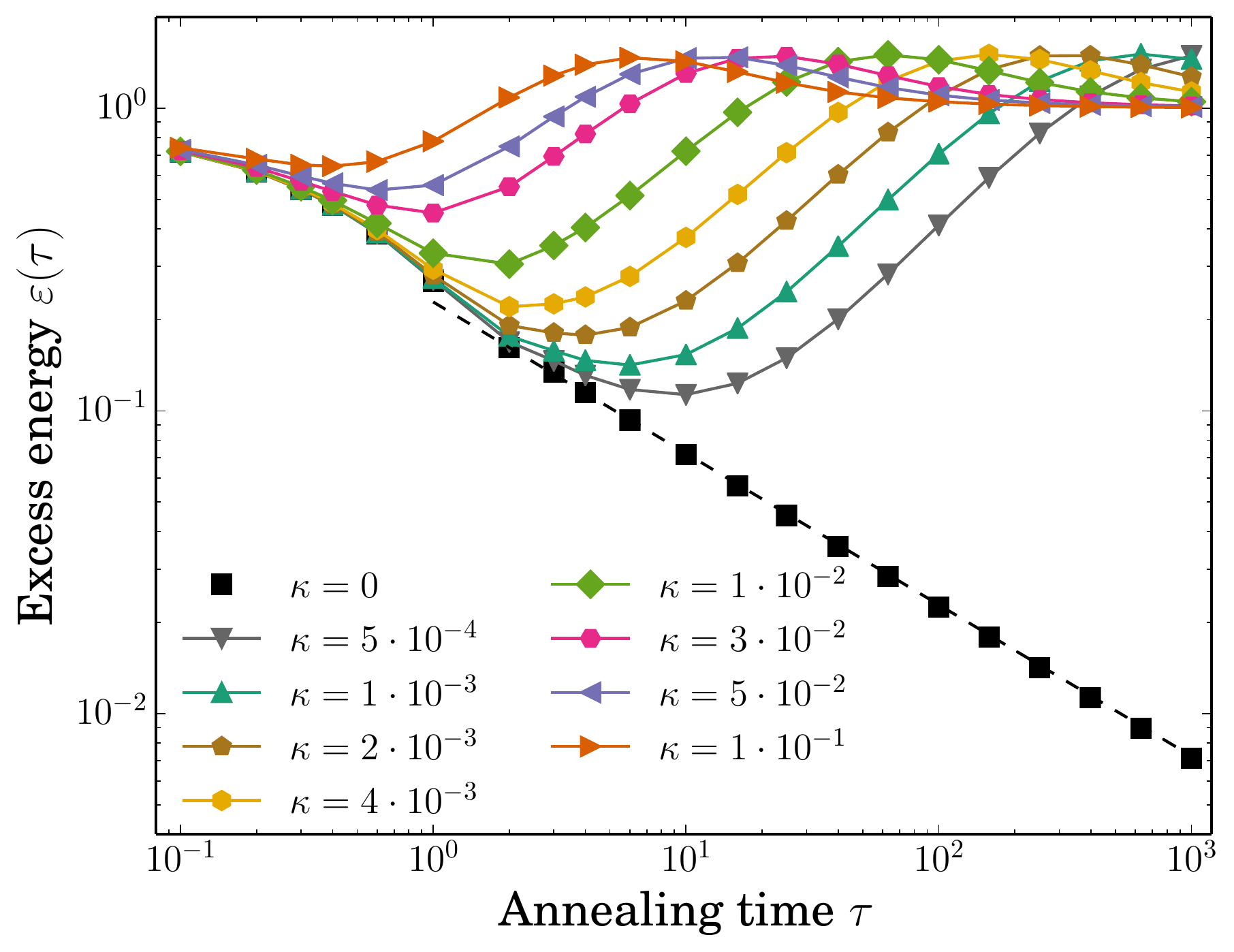}
  \caption{Same plot as in Fig.~\ref{fig:ExcT_FreeF-Pump}, but for a free-fermion model coupled
    to an environment which induces a decay mechanism, as in Eq.~\eqref{eq:L_decay}:
    $L_n^{(2)} = c_n$. We observe the same initial trend as for the pumping mechanism, however for longer annealing times we observe an overshooting before the saturation sets in.}
  \label{fig:ExcT_FreeF-Dec}
\end{figure}

Here we study the interplay between pumping and decaying mechanism and the question whether the steady state
of a system subject to both mechanisms is thermal or not.
For this, first we focus on the annealing protocol in the presence of a uniform incoherent decay mechanism only,
induced by the Lindblad operators $L_n^{(2)} = c_n$. The behaviour of the final excess energy
$\varepsilon(\tau)$ as a function of the annealing time is shown in Fig.~\ref{fig:ExcT_FreeF-Dec}
for different values of the dissipation strength $\kappa$.
As one can see from the figure, at relatively small annealing times
the trend is qualitatively analogous to that obtained for the incoherent pumping
(see Fig.~\ref{fig:ExcT_FreeF-Pump}). The non-monotonic behaviour of $\varepsilon(\tau)$
reveals the presence of an optimal working point, where the number of defects is minimal.
However for larger times $\tau$ we also recognize the appearance
of an overshooting point, where the energy defects become larger than those reached for an infinitely
slow annealing.
Here as well, we have checked that the behaviour of $\varepsilon(\tau) - \varepsilon(+\infty)$,
after such overshooting point, decays with a power law as $\sim \tau^{-1}$,
and again a linear scaling with $\kappa$~\cite{CamposVenuti}.

To better highlight the overshooting behaviour, let us recall that, contrary to the incoherent
pumping mechanism, the incoherent decay will drastically affect the completely filled
ground state of the initial Hamiltonian at $\Gamma(t_{\rm in}) = +\infty$,
since it would tend to empty the system and thereby increasing the energy in the system.
Consequently in the limit $\tau \to \infty$, where we can assume the system always to be in the instantaneous steady state,
its energy will be $E(t) = 2 \Gamma(t) > 0$, so it will approach its final value $E(t_{\rm fin})=0$ from above.
Since for $1 \ll \tau < \infty$ we know that the dynamics approximately follows this open adiabatic dynamics,
it is reasonable to expect that its instantaneous energy will follow a similar trend, in particular it will approach
its final value from above as well, and the corrections due to finite $\tau$ will result in the observed overshooting.

\begin{figure}[!t]
  \centering
  \includegraphics[width=\columnwidth]{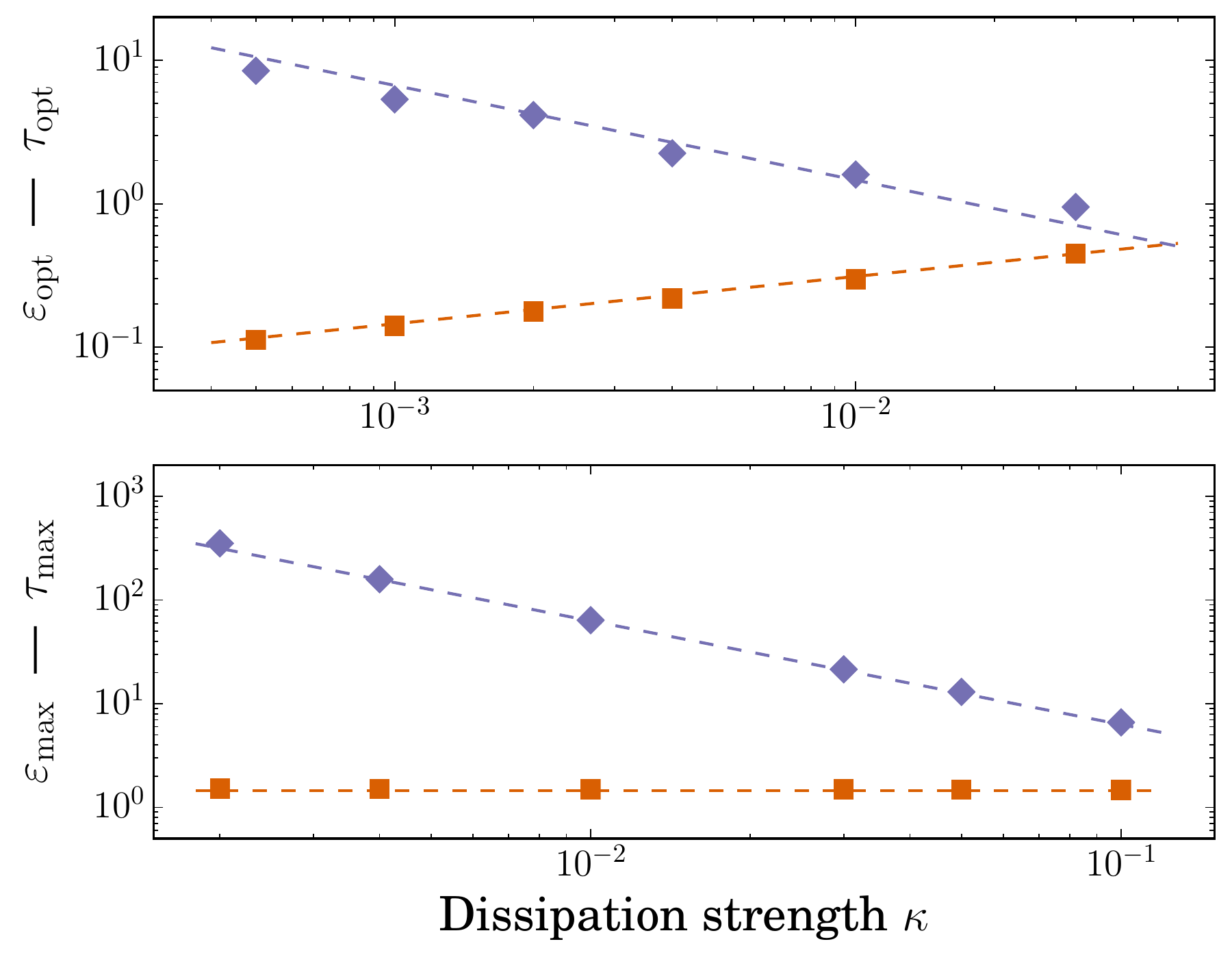}
  \caption{Excess energy (orange squares) and corresponding annealing time (violet diamonds)
    as a function of the dissipation strength $\kappa$, in the presence of an incoherent decay
    mechanism, both for the optimal working point (upper panel) and for the overshooting point (lower panel).
    Numerical data (symbols) are obtained using the same parameters as in Fig.~\ref{fig:ExcT_FreeF-Dec},
    and agree with a power-law behaviour (dashed lines) with slopes
    $1/3$ and $-0.2/3$ (upper), and a constant value as well as a slope of $-1$ (lower).}
  \label{fig:OptT_MaxT_FreeF-Dec}
\end{figure}

In Fig.~\ref{fig:OptT_MaxT_FreeF-Dec} (upper panel) we analysed the minimum excess energy that is reached
at the optimal working point, and the corresponding annealing time.
Their behaviour with $\kappa$ again follows a power law which is similar to the pumping case,
as discussed in Sec.~\ref{sec:ScalOpt}. Note that the argument leading to the scaling predictions for a pumping environment
holds as well for a decaying environment, only the function in Eq.~\eqref{eq:scaling_function} changes.
However, this does not influence the scaling behaviour discussed here, but only the pre-factors.
We also stress that, in the decay case, the integral involved in this calculation strongly depends on the value $-5 \tau$
used to replace the initial value of the field by $-\infty$, which is reasonable since we have seen that
this environment creates defects already long before the quantum critical region is reached.
As a consequence, the scaling behaviour of $\varepsilon_{\rm opt}$ and the corresponding $\tau$ behaves 
in accordance with Eqs.~\eqref{eq:TauOpt}-\eqref{eq:NOpt}.

In the lower panel of Fig.~\ref{fig:OptT_MaxT_FreeF-Dec} we have repeated a similar analysis for the maximum
excess energy at the overshooting point and the corresponding annealing time, as
a function of the dissipation strength.
We observe that such annealing time $\tau_{\rm max}$ scales linearly with $\kappa$,
while the change of the maximum excess energy $\varepsilon_{\rm max}$ is relatively small,
since it varies by less than $10 \%$ over almost two orders of magnitude.

\begin{figure}[!t]
  \centering
  \includegraphics[width=\columnwidth]{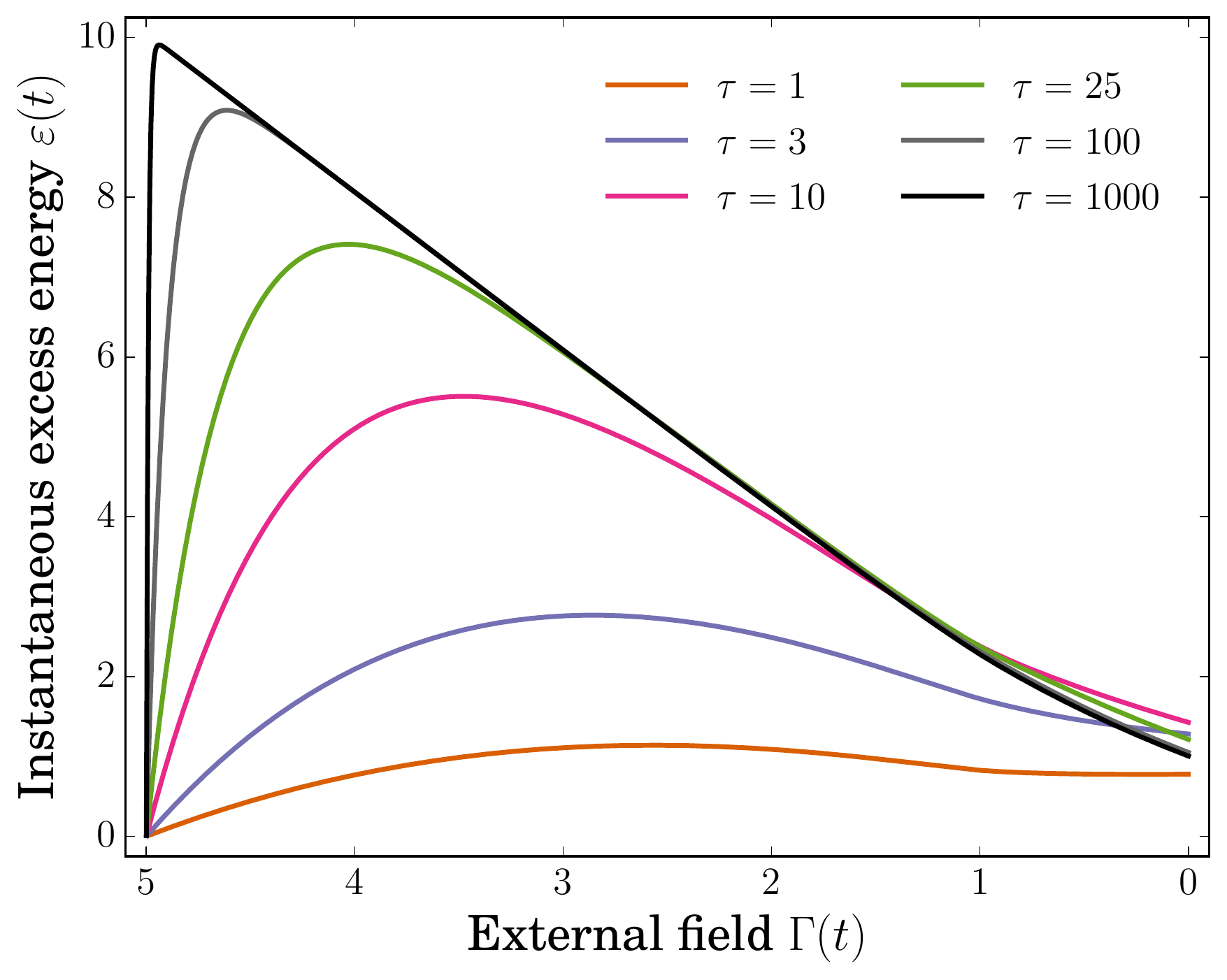}
  \caption{Instantaneous excess energy as a function of the external field $\Gamma(t)$,
    for various annealing times $\tau$ and fixed dissipation strength $\kappa=0.1$ of the decaying environment.
    One observes that for long annealing times ($\tau =1000$) the system follows the instantaneous steady state,
    which has an energy $2 \Gamma(t)$, and at the end it saturates toward $\varepsilon(\infty) = 1$.
    Intermediate times show the same trend but are not following the open adiabatic dynamics as closely,
    thus resulting in a higher final excess energy $\varepsilon(\tau) > \varepsilon(\infty)$.
    For very short annealing times ($\tau = 1$), the influence of the dissipation is much smaller
    and the final excess energy is smaller than $\varepsilon(\infty)$.}
  \label{fig:InExcT_FreeF-Dec}
\end{figure}

For a better understanding of the overshooting, in Fig.~\ref{fig:InExcT_FreeF-Dec} we show the instantaneous excess energies
for different annealing times during the protocol. For very small annealing times we see that the instantaneous
steady-state energy is far away from the actual dynamics and no overshooting takes place.
For long annealing times, the excess energy increases hugely in the beginning and then follows the (open) adiabatic dynamics,
while the behaviour is similar for intermediate annealing times, but not as drastic.
As a consequence, there is an intermediate regime where the annealing time $\tau$ is big enough such that
an overshooting can take place, and the final excess energy $\varepsilon(\tau)$ will be bigger
than in the infinite-time limit $\varepsilon(\infty)$.

\begin{figure}[!t]
  \centering
  \includegraphics[width=\columnwidth]{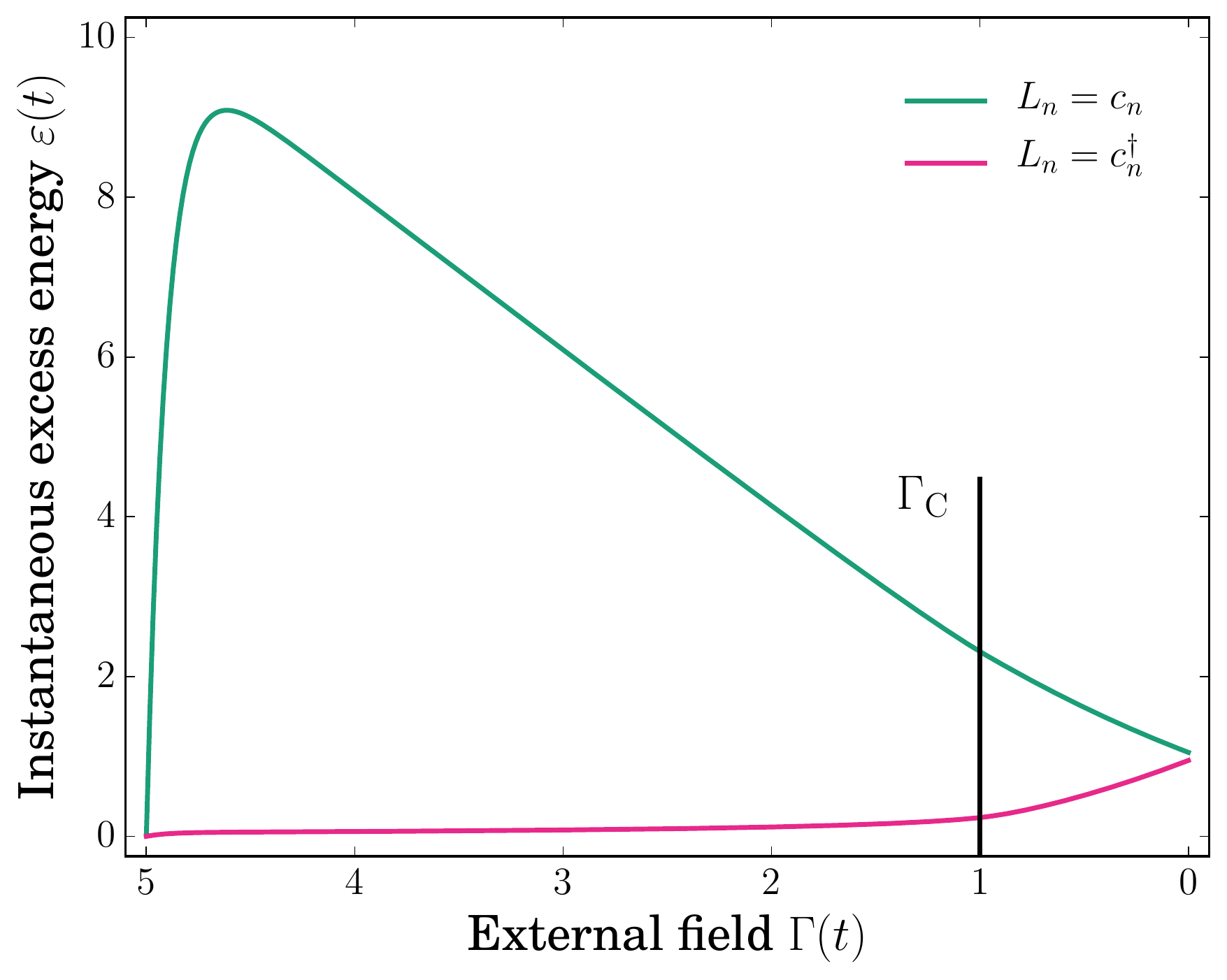}
  \caption{Instantaneous excess energy during the annealing protocol as a function of the external field $\Gamma(t)$
    for the two different types of dissipation. $\Gamma_c=1$ locates the critical point where the Ising-like
    quantum phase transition at zero temperature occurs. Here we fixed $\kappa=0.1$.}
  \label{fig:InExcT_FreeF-DecPump}
\end{figure}

To underline the difference between the two kinds of dissipation (pumping/decay), in Fig.~\ref{fig:InExcT_FreeF-DecPump}
we plotted the instantaneous excess energy for the same parameters ($\tau = 10^3$, $\kappa=10^{-1}$), but different type
of dissipation. We observe that, as stated above, in the pumping case the excess energy mostly increases in the last fifth
of the protocol, which is close to the quantum critical point.
The decaying scenario shows a completely different behaviour, rather following adiabatically
the instantaneous steady state of the system.

\begin{figure}[!t]
  \centering
  \includegraphics[width=\columnwidth]{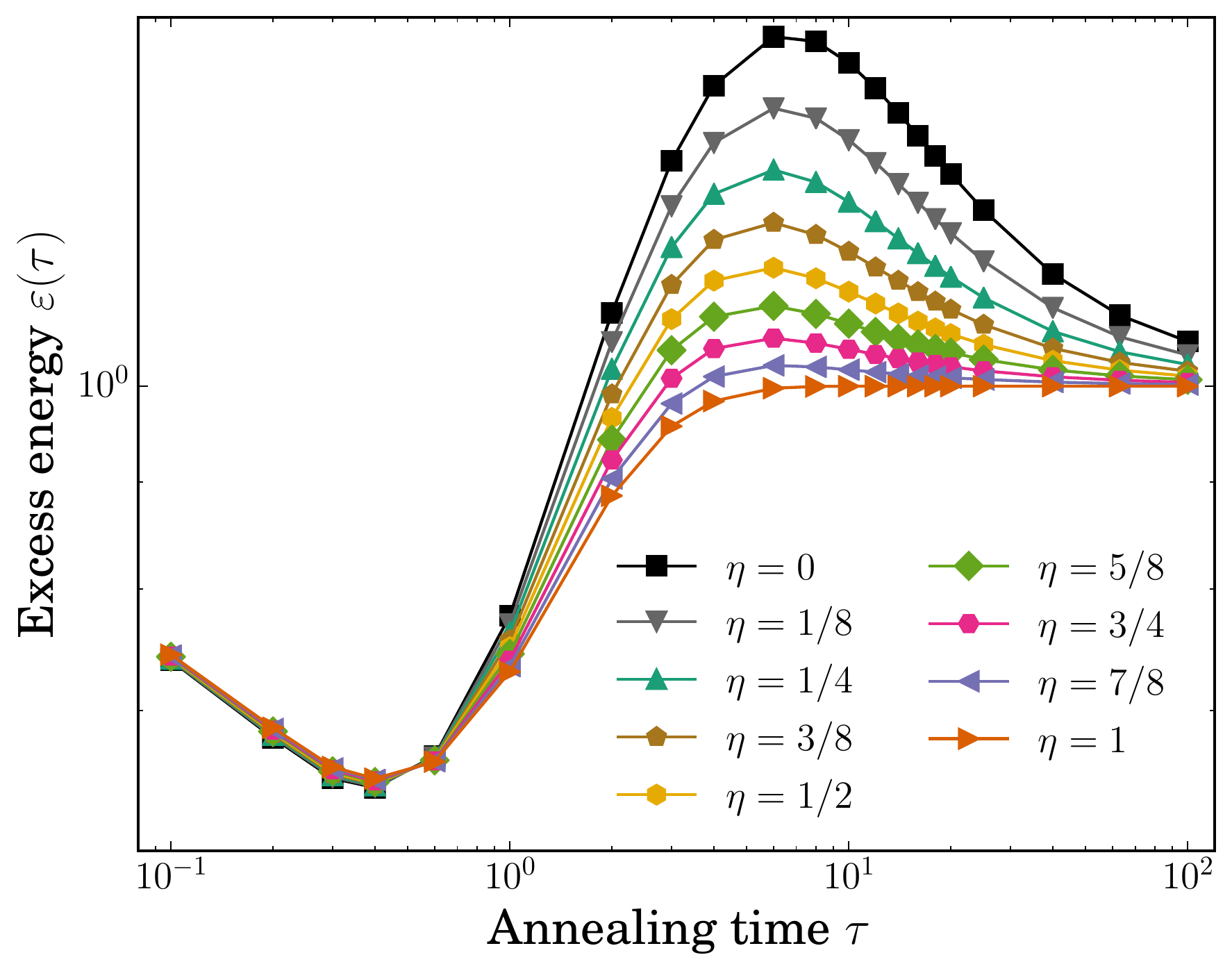}
  \caption{Final excess energy as a function of the annealing time coupled to an environment which induces
    both a pumping as well as a decaying mechanism.
    The various data sets denote different values of the ratio between the two $\eta$ as listed in the legend. Here we simulated the annealing protocol of
    Eq.~\eqref{eq:annealing} for chains of $L=10^3$ sites.}
  \label{fig:ExcT_FreeF-DecPump}
\end{figure}

\begin{figure}[!t]
  \centering
  \includegraphics[width=\columnwidth]{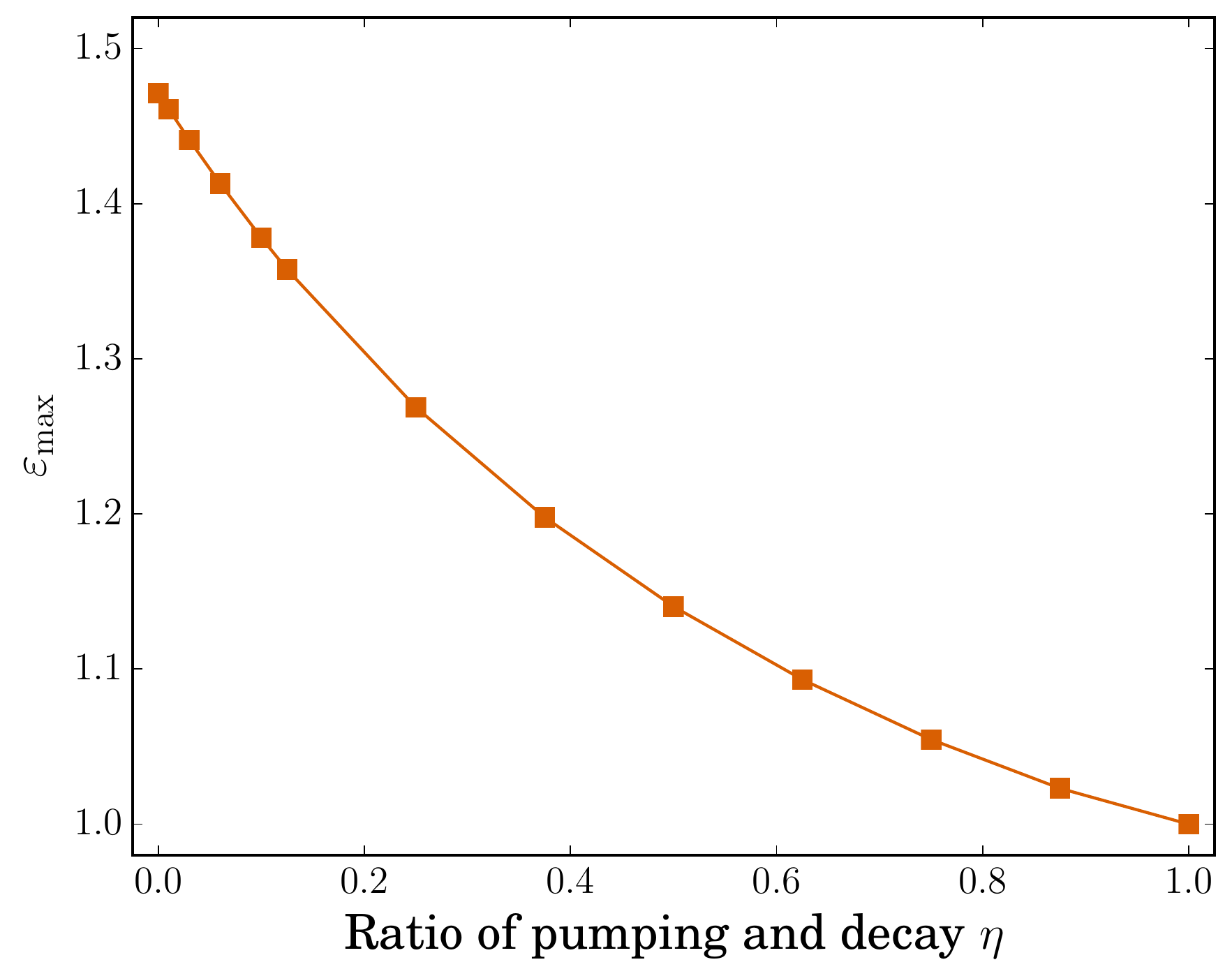}
  \caption{Maximum value of the excess energy at the overshooting point during the annealing protocol,
    as obtained from the data in Fig.~\ref{fig:ExcT_FreeF-DecPump}, as a function of the ratio $\eta$.}
  \label{fig:MaxT_FreeF-DecPump}
\end{figure}

Now, we turn our attention to the interplay between pumping and decay and how the overshooting observed for pure decay
is influenced. For this we study the final excess energy as a function of the ratio between pumping and decaying,
$\eta = \kappa_{\rm pump} / \kappa_{\rm decay}$.
In Fig.~\ref{fig:ExcT_FreeF-DecPump} we show the results for values of $\eta$ ranging from $0$ (no pumping),
where we observe the biggest overshooting, to $1$, where the overshooting is completely disappeared.
Note that the optimal working point does not change much with varying $\eta$ since, for the given parameters,
the contribution by the pumping mechanism in this regime are far smaller than the one by the decay.
The scaling of the maximum value of the overshooting as a function of $\eta$ is shown in Fig.~\ref{fig:MaxT_FreeF-DecPump}.
An explanation for the diminishing overshooting can be given when looking at the dependence of the instantaneous
steady-state energy during the protocol: if $\eta$ is smaller than $1$, it decreases from an initial positive value
to $0$ linearly, such that an overshooting is possible. For $\eta = 1$, the instantaneous steady-state energy
is constant equal to $0$ such that an overshooting due to adiabatic dynamics is prevented. 
For $\eta > 1$, the instantaneous steady-state energy approaches $0$ linearly from below, again preventing an overshooting.

Finally, we comment on the issue of thermalisation: A single qubit subjected to both incoherent pumping
($L^{(1)} = c^{\dagger}$) and decay ($L^{(2)} = c$) processes
would relax to a thermal state whose inverse temperature $\beta$ is related to the ratio
of the strengths of the two Lindblad operators.
Since in our case the translational invariant quadratic Hamiltonian $H(t)$ factorizes
into many Hamiltonians (each one describing a mode of pseudo-momentum $k$) whose Hilbert spaces 
are, each of them, essentially two-dimensional,
this might raise the question if our system shows thermalisation as well.
Indeed, the steady state of each of these modes can be approximated by a thermal state
with very high fidelity ($>98\%$) for the complete range of physical relevant coupling strengths.
However, the corresponding inverse temperature $\beta$ ($k_{\rm B} = 1$) of each mode depends on $k$,
and therefore the complete steady state is not well approximated by a thermal state of a single parameter $\beta$.

\subsection{Dephasing}
\label{sec:Dephasing}

Up to now all the discussion was based on a system-bath coupling scheme which induces
a decay/pumping mechanism. There is however a complementary effect of decoherence,
where the dissipation can generate pure dephasing. This can be easily obtained
through diagonal Lindblad terms $L_n^{(3)} = c^\dagger_n c_n$ (which are proportional
to the onsite fermionic number operator), as in Eq.~\eqref{eq:L_dephas}.
As detailed in App.~\ref{app:TrInv_deph}, despite the translational invariance,
in such case the solution to the master equation~\eqref{eq:master} cannot be trivially written
in a tensor structure as that in Eq.~\eqref{eq:RhoT_k}.
As a matter of fact, the Lindbladian $\mathbb{D}[\rho]$ now transforms into a non-local object,
where the different momentum modes are now coupled together.
Therefore it is more suitable to solve a close set of $4L$ differential linear equations
for the relevant two-point correlators~\cite{Eisler_2011}, see Eq.~\eqref{eq:timedep_homo}.
By employing a fourth-order Runge-Kutta integration procedure of those equations,
with time step $dt = 10^{-2}$, we were able to reach annealing times up to $\tau = 10^3$.

\begin{figure}[t]
  \centering
  \includegraphics[width=\columnwidth]{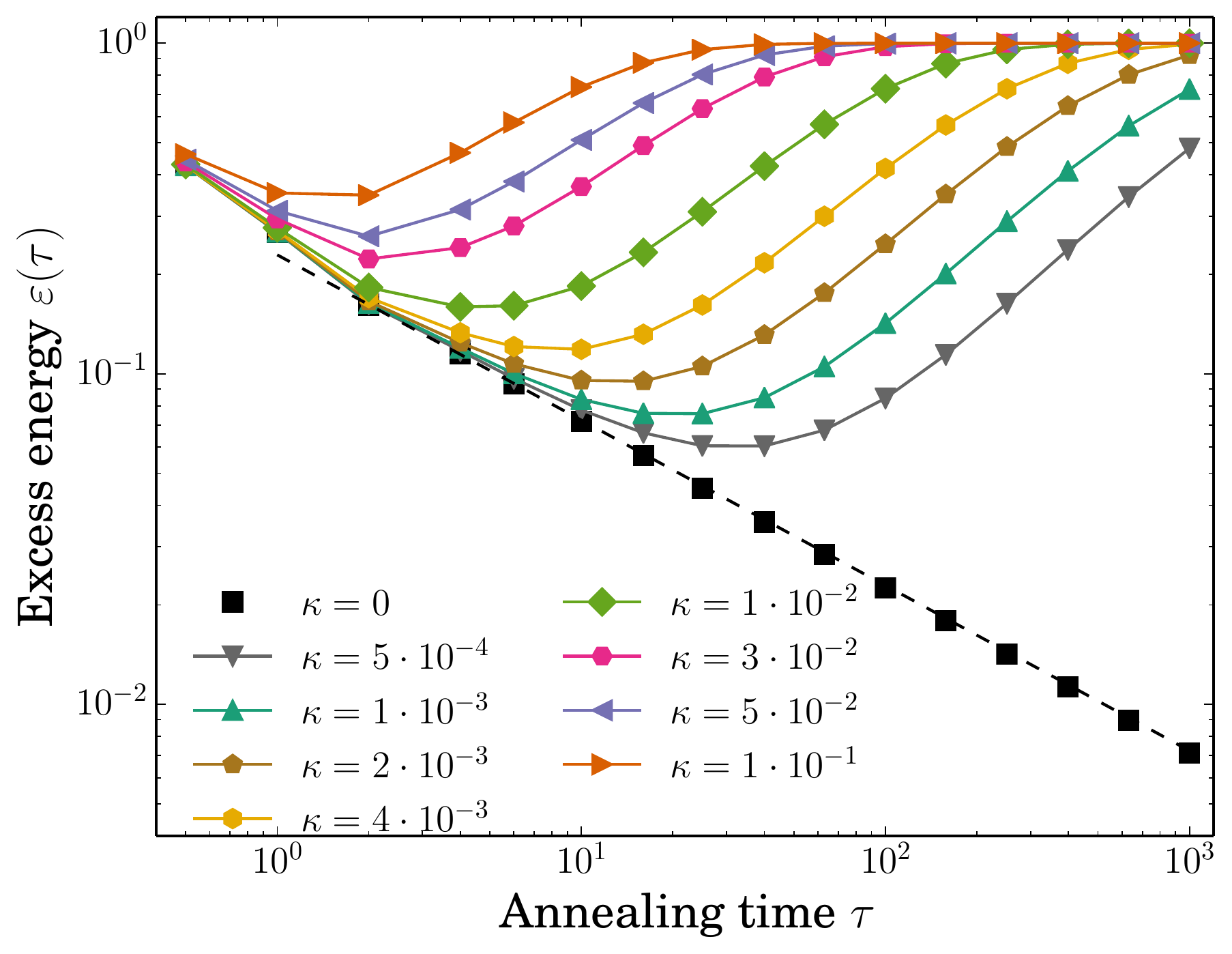}
  \includegraphics[width=\columnwidth]{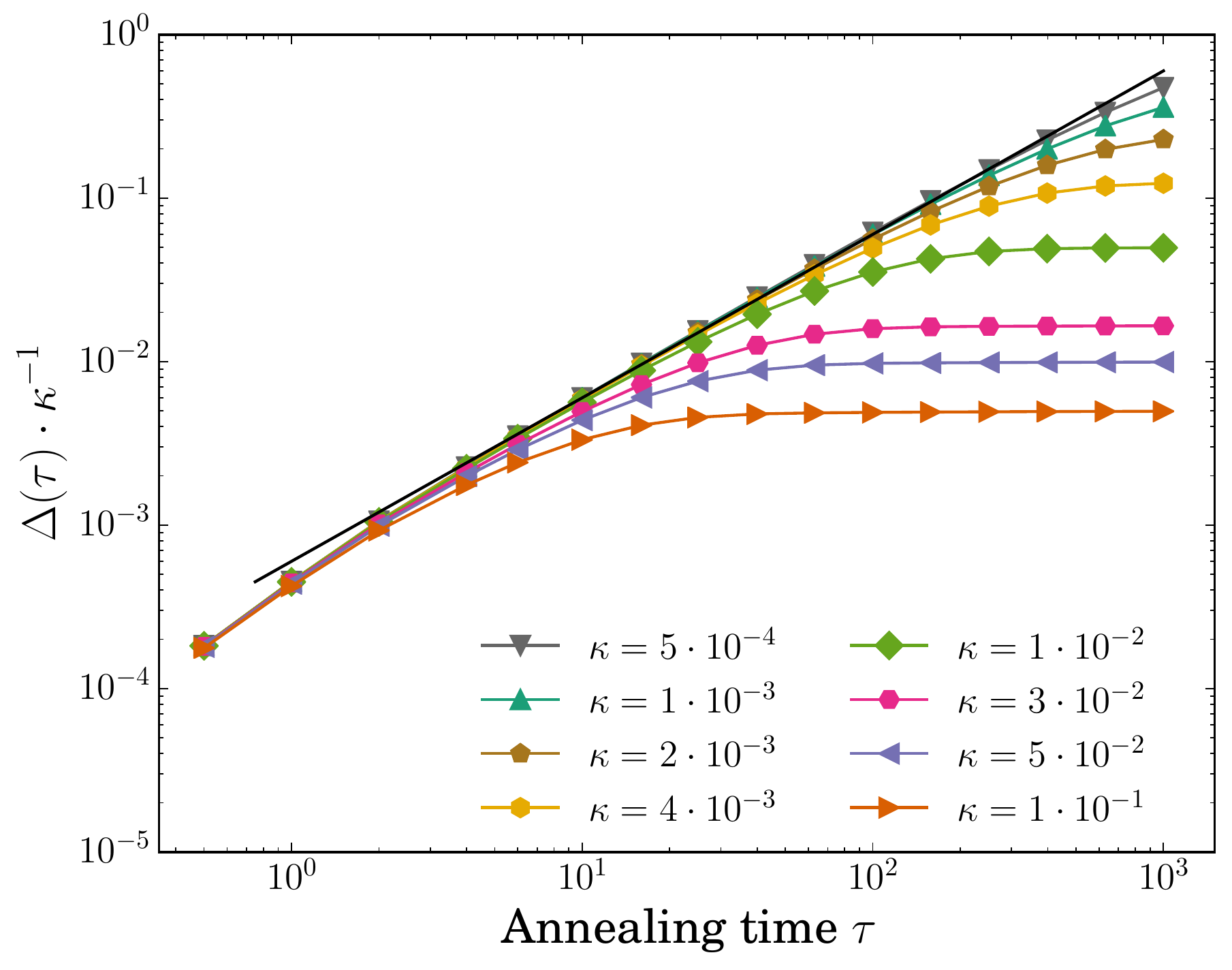}
  \caption{Final excess energy $\varepsilon(\tau)$ (upper panel) and rescaled difference
    $\Delta (\tau) / \kappa$ (lower panel) as a function of the annealing time $\tau$,
    in the free-fermion model~\eqref{eq:Ham_quadratic} coupled to a dephasing
    environment $L^{(3)}_n = c^\dagger_n c_n$. Here we simulated the annealing protocol of
    Eq.~\eqref{eq:annealing} for chains of $L=501$ sites.
    The other parameters are set as in Fig.~\ref{fig:ExcT_FreeF-Pump}.}
  \label{fig:ExcT_RsclK_FreeF-Deph}
\end{figure}

The main results of our analysis are summarized in Fig.~\ref{fig:ExcT_RsclK_FreeF-Deph}, where we plot
(upper panel) the excess energy $\varepsilon(\tau)$ at the end of the annealing,
as a function of the annealing time $\tau$.
Comparing these data with those of Fig.~\ref{fig:ExcT_FreeF-Pump}, we immediately recognize
a qualitatively analogous trend as for the pumping mechanism.
In particular, the non-monotonic behaviour again reveals a competing effect between the KZ mechanism
and the incoherent dephasing.
Quantitative differences are barely visible on the scale of the two figures.
We observed a slight worsening of the annealing protocol,
for the same value of $\tau$, the excess energy being slightly larger than that
of the previous case.
As we did previously, we also analysed the excess-energy difference $\Delta(\tau)$
rescaled by $\kappa$ (bottom panel). Its scaling with $\tau$ is completely analogous
to that in Fig.~\ref{fig:Rscl_FreeF-Pump},
with data growing linearly with the annealing time, and eventually deviating for
sufficiently large values of $\kappa$ and $\tau$.

\begin{figure}[!t]
  \centering
  \includegraphics[width=\columnwidth]{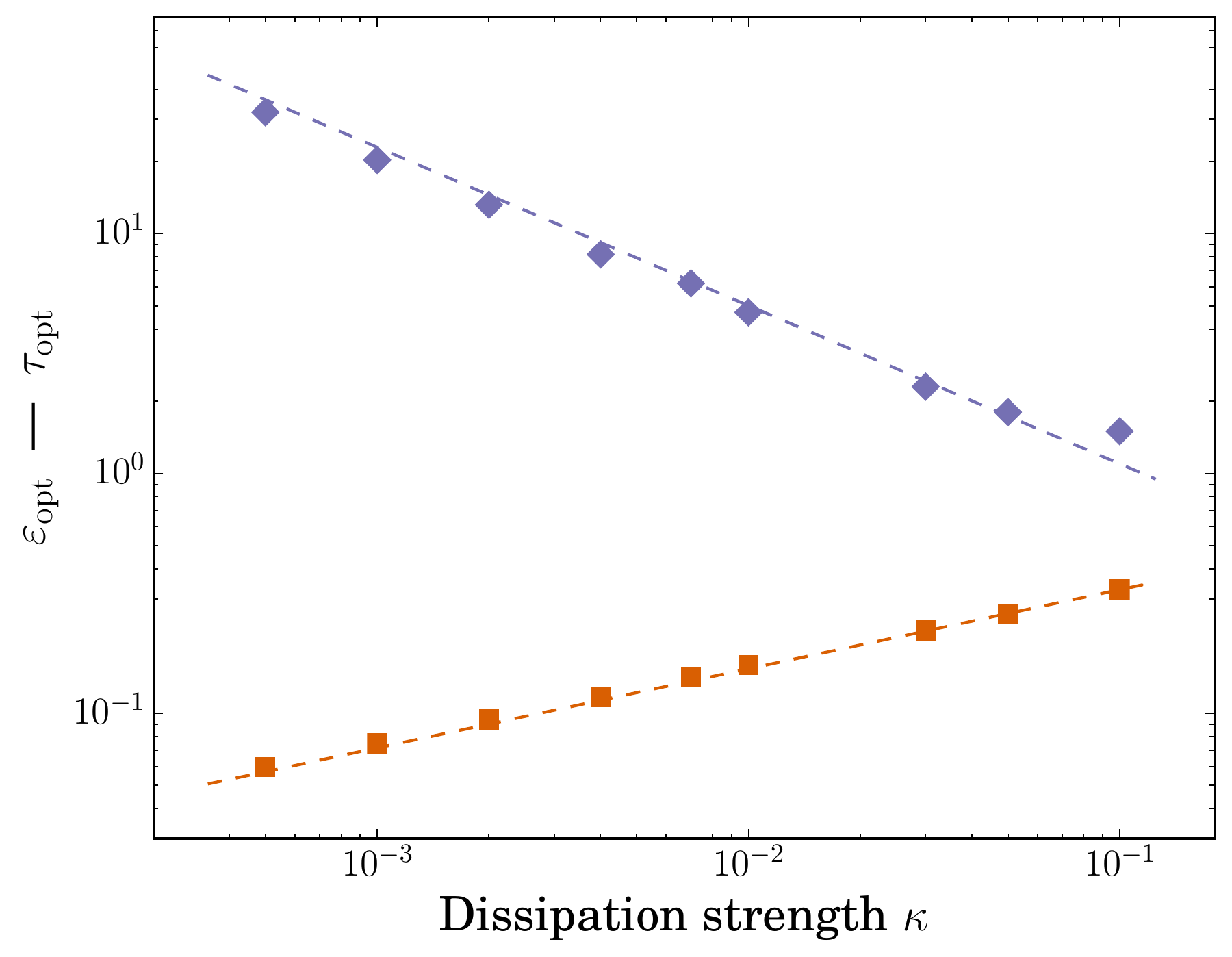}
  \caption{Optimal excess energy (orange squares) and corresponding annealing time (violet diamonds)
    as a function of the dissipation strength.
    Dashed lines denote power-laws with slopes $1/3$ and $-2/3$, respectively
    for $\varepsilon_{\rm opt}$ and for $\tau_{\rm opt}$.
    Data are taken from Fig.~\ref{fig:ExcT_RsclK_FreeF-Deph}, and refer to the free-fermion model
    with a dephasing environment.}
  \label{fig:OptT_FreeF-Deph}
\end{figure}

Finally we recall that the argument of Sec.~\ref{sec:ScalOpt} for determining the scaling
of the optimal working point for the annealing protocol as a function of $\kappa$
holds also in this case. Indeed the corresponding data (with the same power-laws), shown in Fig.~\ref{fig:OptT_FreeF-Deph},
are closely similar to those of Fig.~\ref{fig:OptT_FreeF-Pump}).

Summarizing the results of our analysis on the quantum annealing in a translationally invariant
free-fermion model interacting with a local environment, the emerging scenario for the
different types of dissipation is the following.
For all the three incoherent mechanisms we observe a competition which leads to the onset of
an optimal working point for the annealing procedure at a given $\tau_{\rm opt}$ rate.
On the other side, for larger values of $\tau$ an overshooting point appears only
in the presence of a decay mechanism, due to the fact that the instantaneous energy approaches
the steady-state value $\varepsilon(\tau)=1$ from below (while the opposite happens for
the pumping and for the dephasing).
Finally, we analysed how the final excess energy approaches the $\tau \to \infty$ limit:
while for pumping and decay we observed a behaviour $|\varepsilon(\tau) - \varepsilon(\infty)| \sim \tau^{-1}$,
for dephasing we found $|\varepsilon(\tau)-\varepsilon(\infty)| \sim \exp(-\tau)$.

\section{Ising chain}
\label{sec:Ising}

Let us now go back to the spin-1/2 language and discuss the effects of the coupling
to an external bath on the quantum annealing of the Ising chain, Eq.~\eqref{eq:Ham_Ising}.
We first notice that dephasing can be induced by a Lindblad term $L_n^{(3a)} = \sigma^z_n$,
which is readily mapped into the local fermionic operator $(2 c^\dagger_n c_n -1)$,
through the JWT of Eq.~\eqref{eq:JWT}.
In such case, one would thus recover the dephasing mechanism for free fermions
(we refer to Sec.~\ref{sec:Dephasing} for details).
On the other hand, incoherent pumping/decay would be induced by $L_n^{(1a)} = \sigma^+_n$
and $L_n^{(2a)} = \sigma^-_n$, respectively;
in that case, when mapping into fermions, the appearance of the JW string operator
forbids an analytic treatment as the one discussed previously.
Let us thus concentrate on the latter scenario.

We employ a numerical method based on an efficient approximation of the many-body
density matrix in terms of a MPO~\cite{Verstraete_2004, Zwolak_2004}.
We expect this to be valid whenever the amount of correlations in the system
is sufficiently small to satisfy an area-law scaling for the bipartite entanglement
in the operator space.
The time evolution is performed by means of the time-evolving block decimation
(TEBD) algorithm, after a Trotter decomposition
of the Liouvillian superoperator in the right-hand side of Eq.~\eqref{eq:master}.
In our simulations of the annealing protocol~\eqref{eq:QAnnealing}
for the Ising model we considered systems up to $L=20$ sites,
using MPOs with a bond link $m \approx 250$ and adopting a typical Trotter step $dt=10^{-2}$.
We adopted the same time dependence of the field $\Gamma(t)$ as in Eq.~\eqref{eq:annealing},
where for practical convenience we started from $t_{\rm in} = -3\tau$, and verified that
(on the scales of the figures shown below) the results are not affected by this choice.
As detailed below, we found an emerging physical scenario which is consistent to that 
previously discussed in Sec.~\ref{sec:TraslInv}, already for small sizes $L \gtrsim 10$.

\begin{figure}[t]
  \centering
   \includegraphics[width=\columnwidth]{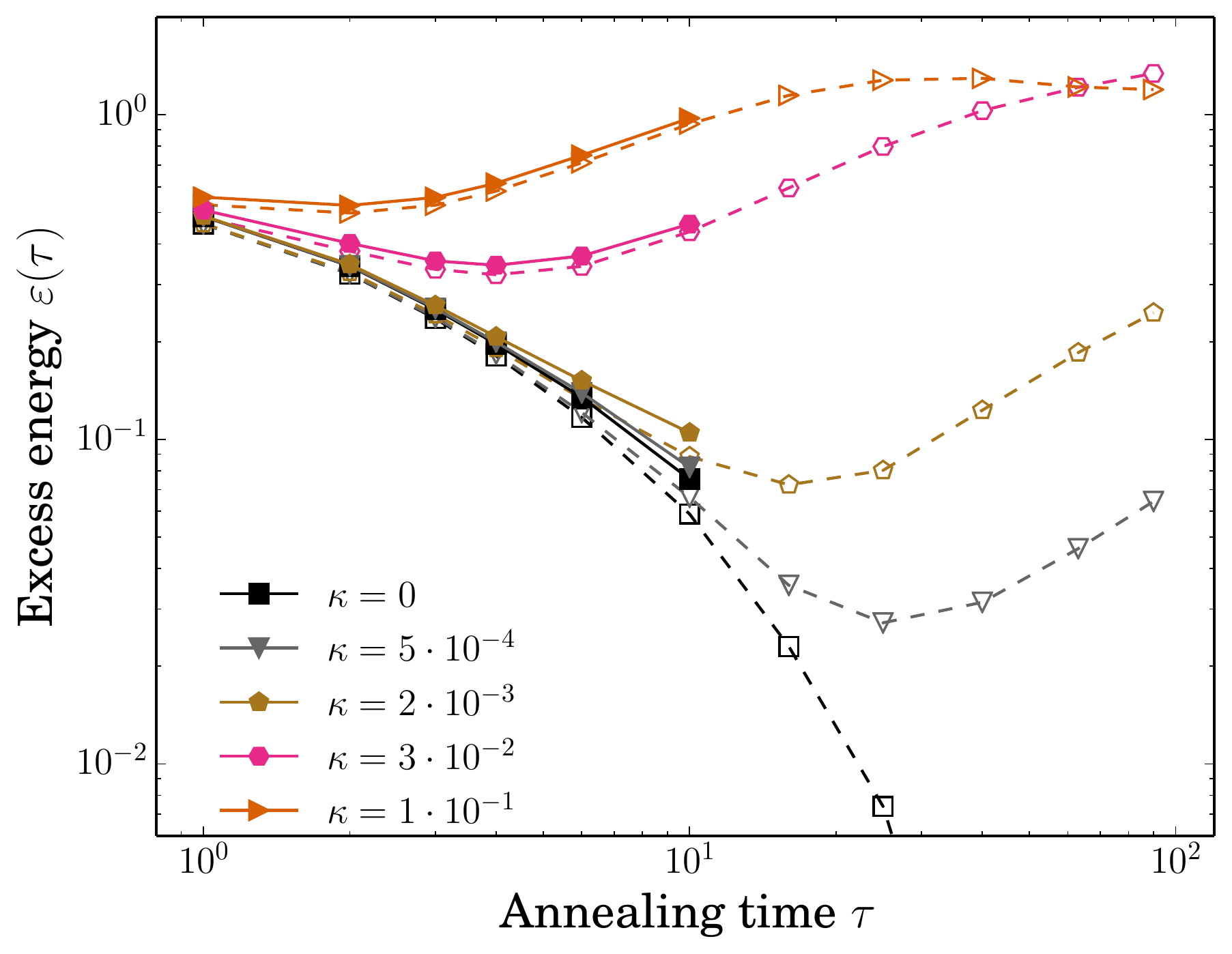}
  \caption{Final excess energy as a function of the annealing time, for the Ising
    chain~\eqref{eq:Ham_Ising} coupled to an environment through Lindblad operators
    inducing a decay mechanism $L_n^{(2a)}= \sigma^-_n$.
    The various data sets denote different values of the dissipative coupling $\kappa$,
    as listed in the legend. Filled symbols and continuous lines refer to chains
    with $L=20$ sites, while empty symbols with dashed lines are for $L=10$.}
  \label{fig:ExcT_Ising}
\end{figure}

\begin{figure}[!t]
  \centering
  \includegraphics[width=\columnwidth]{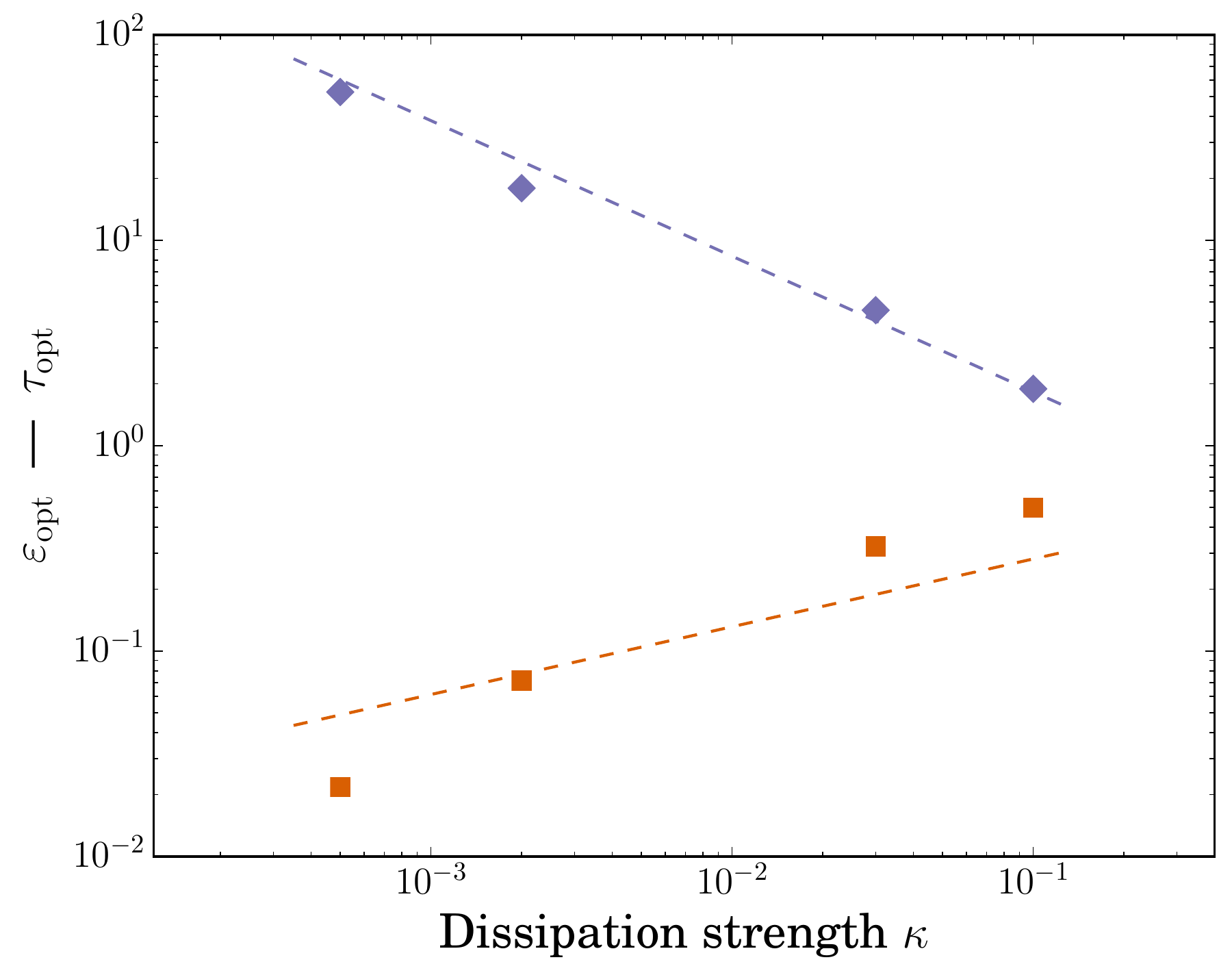}
  \caption{Optimal excess energy (orange squares) and corresponding annealing time (violet diamonds)
    as a function of the dissipation strength.
    Numerical data (symbols) are obtained using the same parameters as in Fig.~\ref{fig:ExcT_Ising},
    for $L=10$, and fairly agree with power laws (dashed lines) of slopes $1/3$ and $-2/3$, respectively.}
  \label{OptT_Ising}
\end{figure}

Our numerical results for the annealing in the presence of incoherent decay,
showing the final excess energy as a function of $\tau$ and for various dissipation strengths $\kappa$,
are summarized in Fig.~\ref{fig:ExcT_Ising}.
Despite the KZ power-law scaling cannot be seen for limited system sizes
(not even in the absence of dissipation), the non-monotonicity of the various curves
for $\kappa \neq 0$ clearly emerges as a result of the open-system dynamics.
We ascribe this behaviour to the emerging picture described in Sec.~\ref{sec:TraslInv},
where we discussed much longer systems of free fermions.
Indeed, in Fig.~\ref{OptT_Ising} we repeated the same analysis for the scaling of the optimal
working time $\tau_{\rm opt}$ and of the corresponding optimal
excess energy $\varepsilon_{\rm opt}$ with the dissipation strength, finding a similar power-law behaviour.
The exponents do agree within $20\%$ of relative difference. 
We point out that we were not able to fully resolve the overshooting behaviour in this case,
since it would require longer annealing times. However this is already visible
in Fig.~\ref{fig:ExcT_Ising}, for the curve corresponding to $\kappa= 0.1$.
Moreover, we also checked that the scaling with $\tau$ of the final excess-energy difference
$\Delta(\tau)$ is again linear for sufficiently small values of $\kappa$ and $\tau$,
as for the fermionic model.

\section{Conclusions}
\label{sec:Conclusions}

We presented an extensive study of the adiabatic dynamics of free-fermion models, being driven
across their quantum critical point, within an open-system approach using local Lindblad operators.
Using the excess energy, we quantified the deviations from adiabatic dynamics of the ground state,
showing a competition between the unitary dynamics following a KZ mechanism and incoherent defect
generation due to dissipation. While being local, the studied environment covers a wide range of possible
sources of dissipation ---varying from decay and pumping, separately or simultaneously, to dephasing---
and at the same time showing a consistent behaviour for all of them:
The competition between the two processes leading to an optimal working point. 
This can be modelled by the Ansatz of independent processes, which brings to a scaling behaviour
that predicts the observed optimal working point in a fairly accurate way.
For larger annealing times, we highlighted the possibility to observe an overshooting point,
where defects become larger than those reached for an infinitely slow annealing.
This effect is intrinsically due to the coupling with an external bath, which drives the system
toward the steady state according to the Liouvillian dynamics of the master equation.

Furthermore, we studied the one dimensional Ising chain, which is closely related to the free-fermion models,
by means of a matrix-product-operator technique, where we found the same behaviour
for small system sizes as well, suggesting a generic nature of the observed phenomena.
Within the framework of free-fermion models, a generalization to higher dimensions is straightforward.

\acknowledgments

We thank S. Barbarino and D. Venturelli for fruitful discussions.
We acknowledge the CINECA award under the ISCRA initiative, for the availability of high performance computing resources and support.
SM gratefully acknowledges the support of the DFG via a Heisenberg fellowship.
This research was partly supported by the EU projects QUIC and RYSQ, by SFB/TRR21 and by CRP-QSYNC.

\appendix

\section{Unitary dynamics}
\label{app:TraslInv_U}

Here we provide technical details concerning the dynamics of the free-fermion model
in Eq.~\eqref{eq:Ham_quadratic}, where periodic boundary conditions are imposed.
Namely,
\begin{equation}
  H(t) = - \sum_{n=1}^L \Big\{ \big( c^\dagger_n c_{n+1} + c^\dagger_n c^\dagger_{n+1} + {\rm H.c.} \big)
  + 2 \Gamma(t) c^\dagger_n c_n \Big\},
  \label{eq:HamTrInv_A}
\end{equation}
with $c_{L+1} = -c_1$ for the positive parity sector,
while $c_{L+1} = c_1$ for the negative parity sector.
Here we have implicitly set the coupling strength to one.

The annealing procedure of Eq.~\eqref{eq:annealing} in this context has been already
studied in Ref.~\cite{Dziarmaga_2005}.
The approach consists in employing a Fourier transform of the type
\begin{equation}
  c_{n} = \dfrac{^{-i\pi/4}}{\sqrt{L}} \sum_{k} c_{k} e^{ikn} ,
  \label{eq:Fourier}
\end{equation}
where the operators $c_{k}^{(\dagger)}$ satisfy canonical anticommutation relations for fermions
as well, and the index $k$ takes values (assuming $L$ to be even, without loss of generality)
$k = \pm 1 \tfrac{\pi}{L}, \pm 3 \tfrac{\pi}{L}, \dots , \pm (L-1) \tfrac{\pi}{L}$.
The resulting Hamiltonian in Fourier space takes the form
\begin{equation}
    H \! = \! \sum_{k} \!\! \Big\{  2c^{\dagger}_{k}c_{k} \big[ -\Gamma(t) - \cos k \big] \!
    + \sin k (c^{\dagger}_{k} c^{\dagger}_{-k} + c_{-k}c_{k}) \! \Big\} .
  \label{eq:HamTraslInv}
\end{equation}
Since $H$ conserves the fermionic parity, the global Hilbert space ${\cal H}$ can be written
as a direct sum over different $k$ subspaces: ${\cal H} = \oplus_{k>0} {\cal H}_k$,
and indeed $H=\sum_{k>0} H_k$ as in Eq.~\eqref{eq:HamTraslInv}.
Each subspace at fixed $k>0$ is built from the two states
$\{ |0\rangle , |1_{k}, 1_{-k}\rangle \}$ or $\{ |1_k\rangle , |1_{-k}\rangle \}$,
depending on the parity number (even or odd, respectively).
The ground state is found in the even parity sector, as one can see from the diagonalization of $H_k$.

The Hamiltonian $H_k$ at a given fixed time $t$, as extrapolated from Eq.~\eqref{eq:HamTraslInv},
can be readily diagonalized by means of a Bogoliubov transformation~\cite{LSM_1961, Young_1997}
\begin{equation}
  c_{k} = u_{k}(t) \gamma_{k} + v^{*}_{-k}(t) \gamma^{\dagger}_{-k} ,
  \label{eq:Bogol}
\end{equation}
so that the ground state is annihilated by all the quasiparticle operators $\gamma_k$.
The system dynamics once the parameter $\Gamma(t)$ is varied can thus be found
by employing the time-dependent Bogoliubov method~\cite{Barouch_1970},
which makes the Ansatz that the instantaneous system wave function $|\psi(t)\rangle$
is annihilated by a set of quasiparticle operators $\tilde \gamma_{k(H)}$,
in the Heisenberg representation, which are defined though the transformation
$c_{k(H)} = u_{k}(t) \tilde \gamma_{k(H)} + v^{*}_{-k}(t) \tilde \gamma^{\dagger}_{-k(H)}$.
This Ansatz satisfies the Heisenberg equation
\begin{equation}
  \tfrac{\mathrm{d}}{\mathrm{d}t} c_{k(H)} = i [ H(t), c_{k(H)} ],
\end{equation}
with the constraint $\tilde \gamma_{k(H)}|\psi(t)\rangle = 0$,
provided the coefficients $u_k(t)$ and $v_k(t)$ obey the time-dependent
Bogoliubov-de Gennes equations
\begin{equation}
  \begin{aligned}
    i \tfrac{\mathrm{d}}{\mathrm{d}t} u_{k} = - 2 u_{k} [\Gamma(t) + \cos k] + 2 v_{k} \sin k ,\\
    i \tfrac{\mathrm{d}}{\mathrm{d}t} v_{k} = + 2 v_{k} [\Gamma(t) + \cos k] + 2 u_{k} \sin k .
  \end{aligned}
\end{equation}
These equations can be integrated starting from the initial condition $\Gamma(-\infty)= + \infty$,
and thus mapping them into a Landau-Zener problem~\cite{Dziarmaga_2005}.

\section{Fermionic decay bath}
\label{app:TraslInv_Decay}

Let us now describe the superimposed action of an environment that induces decay in the system,
so the Lindblad operator on each site $n$ is given by $L_{n} = c_{n}$.
It is important to stress that, when applying the Fourier transform~\eqref{eq:Fourier}
on the dissipative part of the master equation~\eqref{eq:Lindbladian},
this does not mix different modes:
\begin{equation}
  \mathbb{D} [\rho] \equiv \sum_k \mathbb{D}_k [\rho]
  = \kappa \, \Big( \sum_{k} c_k \, \rho \, c^{\dagger}_{k} - \tfrac{1}{2} \{ \rho, c^{\dagger}_{k} c_{k} \} \Big) .
\end{equation}
The reason resides in the fact that each term contains two fermionic operators $c_n^{(\dagger)}$,
such as for example,
$\sum_n c_n \, \rho \, c_n^\dagger \to \sum_n \tfrac{1}{L} \sum_{k,k'} c_k \, \rho \, c_{k'}^\dagger e^{-i(k-k')n}$,
and thus the exponential factor, once summed over $n$, gives a Kronecker delta $\delta_{k,k'}$.

As a consequence, the density matrix factorizes into $\rho(t) = \otimes_{k>0} \rho_{k}(t)$, 
and we can decouple the problem into the same $k$ modes as for the non-dissipative case.
Notice however that the dissipation part violates parity conservation of fermions, therefore
here the different subspaces for a given $k>0$ are built up from the four states
$\{ |0\rangle, |1_{k}\rangle, |1_{-k}\rangle, |1_{k}, 1_{-k}\rangle \}$, 
and not simply from two states.
The Hamiltonian in this basis can be explicitly written as $H = \sum_{k>0} H_{k}$, where
\begin{equation}
  H_{k} = \begin{bmatrix}
    0 & 0 & 0 & 2 \sin k\\
    0 & -2 (\Gamma + \cos k) & 0 & 0\\
    0 & 0 & -2 (\Gamma + \cos k) & 0\\
    2 \sin k & 0 & 0 & -4(\Gamma + \cos k)
  \end{bmatrix},
  \nonumber
\end{equation}
with
\begin{equation}
c_{k} =	\begin{bmatrix}
	0 & 1 & 0 & 0\\
	0 & 0 & 0 & 0\\
	0 & 0 & 0 & 1\\
	0 & 0 & 0 & 0
\end{bmatrix}
\quad \mbox{and} \quad
c_{-k} = \begin{bmatrix}
	0 & 0 & 1 & 0\\
	0 & 0 & 0 & 1\\
	0 & 0 & 0 & 0\\
	0 & 0 & 0 & 0
\end{bmatrix}.
\nonumber
\end{equation}

As a matter of fact, solving the full quantum dynamics of $\rho(t)$ translates into solving
$L/2$ Lindblad equations of dimension 4 for $\rho_k(t)$.
In the vectorized form, they can be written as the following linear differential equations
with dimension 16 ($k>0$):
\begin{align}
  \nonumber
  \tfrac{{\rm d}}{{\rm d}t}  |\rho_k\rangle\rangle = &
  \Big\{ i \big( \mathbb{1} \otimes H_k - H_k \otimes \mathbb{1} \big) \\
  - & \tfrac{\kappa}{2} \Big[ \mathbb{1} \otimes \big( c_k^\dagger c_k + c_{-k}^\dagger c_{-k} \big) 
  + \big( c_k^\dagger c_k + c_{-k}^\dagger c_{-k} \big) \otimes \mathbb{1} \Big] \nonumber \\
  + & \kappa \big( c_k \otimes c_k + c_{-k} \otimes c_{-k} \big)\Big\} |\rho_k\rangle\rangle .
  \label{eq:vectorMEQ}
\end{align}
In practice, for every linear operator $W= \sum_{m,n} W_{mn} |m\rangle \langle n|$ acting on the four-dimensional
Hilbert space ${\cal H}_k$ which is spanned by the basis $\{|m\rangle\}_{m = 1,\ldots,4}$,
we associate a vector in the 16-dimensional
superoperator space ${\cal H}_k \otimes {\cal H}_k$, which is spanned by the basis
$\{ |m\rangle \otimes |n\rangle \}_{m,n=1,\ldots,4}$, using the convention
\begin{equation}
  W_{m,n} \to |W\rangle\rangle_{\phi},\quad \phi = m+(n-1)\, L ,
  \label{eq:vectorization_rho}
\end{equation}
with $|W\rangle\rangle \equiv \sum_{m,n} W_{mn} |m \rangle |n\rangle$.
In this way we have that $|W_1 W_2 W_3 \rangle \rangle = (W_1 \otimes W_3^{\rm T}) |W_2\rangle\rangle$,
where~${}^\text{T}$ denotes the transpose operation.
A vectorization of the master equation for $\rho_k(t)$ using this rule, the fact that
$H_k = H_k^{\rm T} = H_k^\dagger$, and that $c_k^\dagger = c_k^{\rm T}$,
we finally arrive at Eq.~\eqref{eq:vectorMEQ}.
The excess energy~\eqref{eq:excess_energy} is then readily obtained,
since $\rho(t) = \otimes_{k>0} \rho_{k}(t)$.

\section{Fermionic dephasing bath}
\label{app:TrInv_deph}

In the case of dephasing Lindblad operators $L_{n} = c^{\dagger}_{n}c_{n}$ on each site,
the Fourier transform applied to $\mathbb{D}[\rho]$ turns out to yield a non-local object,
since each term now contains four fermionic operators,
and thus it is not possible to decouple the different $k$ modes.
Therefore, this kind of dissipation scheme cannot be directly embedded into the Dziarmaga
formalism~\cite{Dziarmaga_2005} described above.

In the following, it is more convenient to reduce our study to two-point correlators,
since all the relevant quantities for our purposes (such as the excess energy~\eqref{eq:excess_energy},
can be expressed in terms of those correlators. This drastically simplifies the analysis
into a closed set of differential equations which scale linearly (or at most quadratically, for
the non-homogeneous case) with $L$~\cite{Eisler_2011}.
We define
\begin{equation}
  \begin{aligned}
    F_{m,n} \equiv \langle c^{\dagger}_{m}c_{n} \rangle, & \qquad
    G_{m,n} \equiv \langle c_{m}c^{\dagger}_{n} \rangle , \\
    I_{m,n} \equiv \langle c^{\dagger}_{m}c^{\dagger}_{n} \rangle,  & \qquad
    K_{m,n} \equiv \langle c_{m}c_{n} \rangle.
  \end{aligned}
  \label{eq:correlFGIK}
\end{equation}
Using anticommutation relations for fermions and the fact that $(c_m c_n)^\dagger = c_n^\dagger c_m^\dagger$,
we have $G_{m,n} = \delta_{m,n} - F_{n,m}$ and also $K_{m,n}^* = I_{n,m}$.

Here we adopt the Heisenberg representation, where the dynamics is described by means of
an adjoint Lindblad master equation for a given observable $O$:
\begin{eqnarray}
  \label{eq:adj_Master_eq}
  \frac{\mathrm{d}}{\mathrm{d}t} O & = & i [H,O] + \tilde{\mathbb D}[O], \quad \mbox{ where} \\
    \tilde{\mathbb D}[O] & = &
    \frac{1}{2} \sum_{n=1}^L \kappa_n \Big( L^{\dagger}_{n} \, [O,L_{n}] - [O,L^{\dagger}_{n}] \, L_{n} \Big) .
    \label{eq:adj_Diss}
\end{eqnarray}
Since in this appendix we deal with the homogeneous cases, we set $\kappa_n=\kappa$.

We first note that
\begin{equation}
  [c_{m},H] = 2 \sum_{n=1}^{L} A_{m,n} c_{n} + B_{m,n} c^{\dagger}_{n} ,
  \label{eq:c_H}
\end{equation}
where the matrices $A$ and $B$ have been defined in Eq.~\eqref{eq:Ham_quadratic},
with $J_n = h_n = 1$.
Moreover, specializing to the dephasing bath $L_{n} = c^{\dagger}_{n} c_{n}$
with uniform couplings $\kappa_n = \kappa$, we have
\begin{equation}
\begin{aligned}
  \tilde{\mathbb D}[c^{\dagger}_{m}c_{n}] &= - \kappa \, c^{\dagger}_{m}c_{n} (1-\delta_{m,n}),\\
  \tilde{\mathbb D}[c_{m}c^{\dagger}_{n}] &= - \kappa \, c_{m}c^{\dagger}_{n} (1-\delta_{m,n}),\\
  \tilde{\mathbb D}[c^{\dagger}_{m}c^{\dagger}_{n}] &= - \kappa \, c^{\dagger}_{m}c^{\dagger}_{n} (1+\delta_{m,n}),\\
  \tilde{\mathbb D}[c_{m}c_{n}] &= - \kappa \, c_{m}c_{n} (1+\delta_{m,n}).
\end{aligned}
\label{eq:Diss_dephasing}
\end{equation}
The adjoint Lindblad master equation~\eqref{eq:adj_Master_eq} for the Hamiltonian~\eqref{eq:HamTrInv_A}
and the dephasing bath, referred to the operator $c^\dagger_m c_n$, reads:
\begin{align}
  \dfrac{\mathrm{d}}{\mathrm{d}t} c^{\dagger}_{m}c_{n} & =
  i [H,c^{\dagger}_{m}]c_{n} + i c^{\dagger}_{m}[H,c_{n}] + \tilde{\mathbb D}[c^{\dagger}_{m}c_{n}] \nonumber \\
  & = 2i \sum_{j} \big( A_{m,j}c^{\dagger}_{j}c_{n} + B_{m,j}c_{j}c_{n} \nonumber \\
  &- \! A_{n,j} c^{\dagger}_{m}c_{j} \! - \! B_{n,j} c^{\dagger}_{m}c^{\dagger}_{j} \big) \!
  - \! \kappa c^{\dagger}_{m}c_{n} (1 \! - \! \delta_{m,n})
  \label{eq:derF}
\end{align}
and correspondingly, for the other two-point operators,
\begin{align}
  \dfrac{\mathrm{d}}{\mathrm{d}t} c_{m}c^{\dagger}_{n} & =
  2i \sum_{j} \big( -A_{m,j}c_{j}c^{\dagger}_{n} - B_{m,j}c^{\dagger}_{j}c^{\dagger}_{n} \nonumber \\
  & + \! A_{n,j} c_{m}c^{\dagger}_{j} \! + \! B_{n,j} c_{m}c_{j} \big) \!
  - \! \kappa c_{m}c ^{\dagger}_{n} (1 \! - \! \delta_{m,n}) , \! \label{eq:derG} \\
  \dfrac{\mathrm{d}}{\mathrm{d}t} c^{\dagger}_{m}c^{\dagger}_{n} & =
  2i \sum_{j} \big( A_{m,j}c^{\dagger}_{j}c^{\dagger}_{n} + B_{m,j}c_{j}c^{\dagger}_{n} \nonumber \\ 
  & + \! A_{n,j} c^{\dagger}_{m}c^{\dagger}_{j} \! + \! B_{n,j} c^{\dagger}_{m}c_{j} \big) \!
  - \! \kappa c^{\dagger}_{m} c^{\dagger}_{n} (1 \! + \! \delta_{m,n}) , \! \label{eq:derI} \\
  \dfrac{\mathrm{d}}{\mathrm{d}t} c_{m}c_{n} & =
  2i \sum_{j} \big( -A_{m,j}c_{j}c_{n} - B_{m,j}c^{\dagger}_{j}c_{n} \nonumber \\ 
  & - \! A_{n,j} c_{m}c_{j} \! - \! B_{n,j} c_{m}c^{\dagger}_{j} \big) \!
  - \! \kappa c_{m} c_{n} (1 \! + \! \delta_{m,n}) . \! \label{eq:derK}
\end{align}
If we now set $l=n-m, \ l \in [0,L-1]$, for a translational invariant system we can define
$\vec {\cal F}$ such that ${\cal F}_{l=n-m} \equiv F_{m,n}$
(and analogously for $\vec {\cal G}$, $\vec {\cal I}$, $\vec {\cal K}$), in such a way
that Eq.~\eqref{eq:derF} can be rewritten as
\begin{align}
  \dfrac{\mathrm{d}}{\mathrm{d}t} {\cal F}_{l} =
  & 2i \sum_{j} \big( A_{m,j} \, {\cal F}_{l+m-j} + B_{m,j} \, {\cal K}_{l+m-j} \nonumber \\
  & - A_{m+l,j} \, {\cal F}_{j-m} - B_{l+m,j} \, {\cal I}_{j-m} \big) \! - \! \kappa P_l {\cal F}_l ,
  \label{eq:derF2}
\end{align}
where $P_l = 1-\delta_{l,1}$ ($l=1, \ldots L$) are the $L$ components of the vector $\vec P$.
The first term on the right-hand side can be manipulated as
\begin{align}
  \sum_{j} A_{m,j} {\cal F}_{l+m-j}
  &= A_{m,m} {\cal F}_{l} \! + \! A_{m,m+1} {\cal F}_{l-1} \! + \! A_{m,m-1} {\cal F}_{l+1} \nonumber \\
  &= A_{l,l} {\cal F}_{l} + A_{l,l-1} {\cal F}_{l-1} + A_{l,l+1} {\cal F}_{l+1} \nonumber \\
  &= \sum_{j} A_{l,j} {\cal F}_{j} = \big( A \cdot \vec{\cal F} \big)_{l},
\end{align}
where in the second line we used the fact that $A^{\rm T}=A$; moreover, due to translational
invariance of the model, it is possible to shift both indices of $A$ together.
Proceeding in an analogous way, for the other terms of~\eqref{eq:derF2} we find
\begin{align}
  \sum_{j} B_{m,j} {\cal K}_{l+m-j} &= - \big( B \cdot \vec{\cal K} \big)_{l},\\
  \sum_{j} A_{l+m,j} {\cal F}_{j-m} &= \big( A \cdot \vec{\cal F} \big)_{l},\\
  \sum_{j} B_{l+m,j} {\cal I}_{j-m} &= \big( B \cdot \vec{\cal I} \big)_{l}.
\end{align}

It is possible to follow the same type of calculations for the other two-point correlators,
Eqs.~\eqref{eq:derG}-\eqref{eq:derK}, such that the dynamics of all the two-point correlators
defined above can be written in a compact way as a set of time-dependent linear equations
\begin{equation}
  \dfrac{\mathrm{d}}{\mathrm{d}t}
  \begin{bmatrix} \vec{\cal F}\\ \vec{\cal G}\\ \vec{\cal I}\\ \vec{\cal K}\\ \end{bmatrix}
  = \left[ M(t) - \kappa \; {\rm diag} \big( \vec P, \vec P, \vec Q, \vec Q \big) \right] 
  \begin{bmatrix} \vec{\cal F}\\ \vec{\cal G}\\ \vec{\cal I}\\ \vec{\cal K}\\ \end{bmatrix} ,
  \label{eq:timedep_homo}
\end{equation}
where the $4L \times 4L$ matrix $M$ is given by 
\begin{equation}
  M(t) = 2i \begin{bmatrix}
    0 & \phantom{-} 0 &            -B     &            -B\\
    0 & \phantom{-} 0 & \phantom{-} B     & \phantom{-} B\\
    B &            -B & \phantom{-} 2A(t) & \phantom{-} 0\\
    B &            -B & \phantom{-} 0     &         -2A(t)
  \end{bmatrix}
\end{equation}
and the vector $\vec P$ has been defined above, while the $L$ components of
the vector $\vec Q$ are given by $Q_l = 1+\delta_{l,1}$ (with $l = 1, \ldots L$).

Eventually, the instantaneous energy $E(t) \equiv \langle H(t) \rangle$ can be calculated
from the quadratic Hamiltonian~\eqref{eq:HamTrInv_A}:
\begin{align}
  E &= - \! \sum_{n=1}^{L} (F_{n,n+1} \! - \! G_{n,n+1} \! - \! K_{n,n+1}
  \! + \! I_{n,n+1}) \! - \! 2 \Gamma \, F_{n,n} \nonumber \\
  &= - L ({\cal F}_{1} - {\cal G}_{1} - {\cal K}_{1} + {\cal I}_{1}) - 2L \, \Gamma \, {\cal F}_{0},
  \label{eq:energy_homo}
\end{align}
where we used anti-commutation relations to express all terms such that the
resulting index of the translational invariant correlators is non-negative.
To match boundary conditions and parity considerations, we use $L$ odd and are therefore in the negative parity sector.

The initial conditions, for a given Hamiltonian $H(t)$ (that is a certain value of $\Gamma$)
can be immediately found by a Bogoliubov transformation that generalizes Eq.~\eqref{eq:Bogol}
to non-homogeneous quadratic systems~\cite{Dziarmaga_2006}:
\begin{equation}
  c_{i} = \sum_{\mu=1}^{L} \big( U_{i,\mu} \gamma_{\mu} + V^{*}_{i,\mu} \gamma^{\dagger}_{\mu} \big) . 
\end{equation}
The transformation satisfies the properties
$\langle \gamma_{\mu} \gamma^{\dagger}_{\nu} \rangle_{0} = \delta_{\mu,\nu}$ and
$\langle \gamma_{\mu} \gamma_{\nu} \rangle_{0} = \langle \gamma^{\dagger}_{\mu} \gamma^{\dagger}_{\nu} \rangle_{0}
= \langle \gamma^{\dagger}_{\mu} \gamma_{\nu} \rangle_{0} = 0$,
where $\langle \dots \rangle_{0}$ indicates the expectation value over the ground state
of $H(t_{\rm in})$, that is with $\Gamma = +\infty$. This yields
\begin{equation}
  \begin{aligned}
    F_{m,n} (t_{-\infty}) &= \langle c^{\dagger}_{m} c_{n} \rangle_{0} = \left( V V^{\dagger} \right)_{m,n}, \\
    G_{m,n} (t_{-\infty}) &= \langle c_{m} c^{\dagger}_{n}\rangle_{0} = \left( U U^{\dagger} \right)_{m,n},\\
    I_{m,n} (t_{-\infty}) &= \langle c^{\dagger}_{m} c^{\dagger}_{n}\rangle_{0} = \left( V U^{\dagger} \right)_{m,n},\\
    K_{m,n} (t_{-\infty}) &= \langle c_{m} c_{n}\rangle_{0} = \left( U V^{\dagger} \right)_{m,n}.
  \end{aligned}
  \label{eq:CondIn}
\end{equation}
From these equations, exploiting the translational invariance of the system,
we can choose the initial conditions of the system~\eqref{eq:timedep_homo}
by selecting the first column of each of those four matrices.

\end{document}